\documentclass[a4paper,fleqn]{cas-dc}

\usepackage[authoryear,longnamesfirst]{natbib}
\usepackage{natbib}
\usepackage{subcaption}
\usepackage{graphicx}

\usepackage{algorithm2e}

\def\tsc#1{\csdef{#1}{\textsc{\lowercase{#1}}\xspace}}
\tsc{WGM}
\tsc{QE}
\tsc{EP}
\tsc{PMS}
\tsc{BEC}
\tsc{DE}

\begin{document}
\let\WriteBookmarks\relax
\def\floatpagepagefraction{1}
\def\textpagefraction{.001}

\shorttitle{Towards Privacy-Preserving Split Learning}

\shortauthors{G. Higgins et~al.}

\title [mode = title]{Towards Privacy-Preserving Split Learning: Destabilizing Adversarial Inference and Reconstruction Attacks in the Cloud}            
\author[1]{Griffin~Higgins}
\author[1]{Roozbeh~Razavi-Far}
\fnmark[1]
\ead{roozbeh.razavifar@unb.ca}
\cormark[1]
\cortext[cor1]{Corresponding author}
\author[2]{Xichen~Zhang}
\author[1]{Amir~David}
\author[1]{Ali~Ghorbani}
\author[3]{Tongyu~Ge}

\affiliation[1]{organization={Canadian Institute for Cybersecurity, University of New Brunswick},
    addressline={46 Dineen Drive}, 
    city={Fredericton},
    postcode={E3B 5A3}, 
    state={New Brunswick},
    country={Canada}}

\affiliation[2]{organization={Sobey School of Business, Saint Mary’s University},
    city={Halifax},
    postcode={B3H 3C3}, 
    state={Nova Scotia},
    country={Canada}}

\affiliation[3]{organization={Huawei Technologies Canada},
    addressline={300 Hagey Blvd}, 
    city={Waterloo},
    postcode={N2L 0A4}, 
    state={Ontario},
    country={Canada}}

\begin{abstract}
  This work aims to provide both privacy and utility within a split learning framework while considering both forward attribute inference and backward reconstruction attacks. To address this, a novel approach has been proposed, which makes use of class activation maps and autoencoders as a plug-in strategy aiming to increase the user's privacy and destabilize an adversary. The proposed approach is compared with a dimensionality-reduction-based plug-in strategy, which makes use of principal component analysis to transform the feature map onto a lower-dimensional feature space. Our work shows that our proposed autoencoder-based approach is preferred as it can provide protection at an earlier split position over the tested architectures in our setting, and, hence, better utility for resource-constrained devices in edge-cloud collaborative inference ($\mathcal{EC}$) systems.
\end{abstract}

\begin{highlights}
\item The end-to-end pipeline in the $\mathcal{EC}$ system is implemented, which includes the primary image classification with edge-cloud collaboration architecture, the image reconstruction models based on white-box attack, and the sensitive attribute inference attack. The basic components can support comprehensive model evaluations and results comparison.
\item An effective plug-in approach is developed and designed for privacy protection in the $\mathcal{EC}$ system. This novel technique is the combination of AE and CAMs, which can provide explainable and attribute-specific privacy protection with all the plug-in properties, without having to retrain or modify the original $\mathcal{EC}$ model. Training of the plug-in protection is allowed. 
\item A comprehensive set of experiments are conducted to validate the performance of the proposed solutions. The superiority of the proposed novel method with respect to the primary task's performance, image reconstruction, and sensitive attribute inference. Under certain experimental settings, upon applying the proposed protection technology, performance loss at the cloud-side model is around 4\%, whereas the performance drop at the adversarial side is more than 20\%. 
\end{highlights}


\maketitle

\section{Introduction}

The last decade has witnessed unprecedented growth in the number of smart and Internet of Things (IoT) devices, such as mobile phones, tablets, smartwatches, GPS-enabled devices, smart sensors, AI-driven smart assists, edge devices, and other wearable devices \citep{dhar2021survey}.  In addition, Artificial Intelligence (AI) techniques have achieved remarkable progress in various real-world applications, such as natural language understanding and translation \citep{liu2023survey}, image classification \citep{li2021survey}, object detection and segmentation \citep{zhao2019object}, and video processing \citep{jiao2021new}. The extensive advancement of AI technologies resulted in the fast proliferation of on-device AI services.  It has never been easier than today to use AI-driven models on our mobile devices for data labeling, image processing,  business analytics, and decision making \citep{cao2021toward, muhammad2020deep, li2016deep}. With on-device AI models lots of tasks can be completed in a real-time, low latency, and even offline manner. Typical tasks include barcode scanning, face detection, image classification, object detection and tracking, language translation, selfie segmentation, and text recognition. 

Even though the on-device AI applications have bright and broad marketing prospects, they still face critical challenges. The state-of-the-art AI models, such as Visual Geometry Group (VGG) \citep{simonyan2014very}, Residual Network (ResNet) \citep{he2016deep}, and transformers \citep{vaswani2017attention} are not only well known for their superb performance but their huge size and sophisticated architecture as well. For instance, the popular ResNet50 and VGG16 models have 23 million and 138 million parameters, respectively \citep{luo2022computer}.  As for the most famous GPT3 model, its number of parameters is staggering, as high as 175 billion  \citep{gao2023chat}. Therefore, as one major issue, it is infeasible to deploy huge and complex AI models on edge devices due to their computational and storage limitations. 

\begin{figure*}[ht]
    \centering 
    \includegraphics[width=0.95\textwidth]{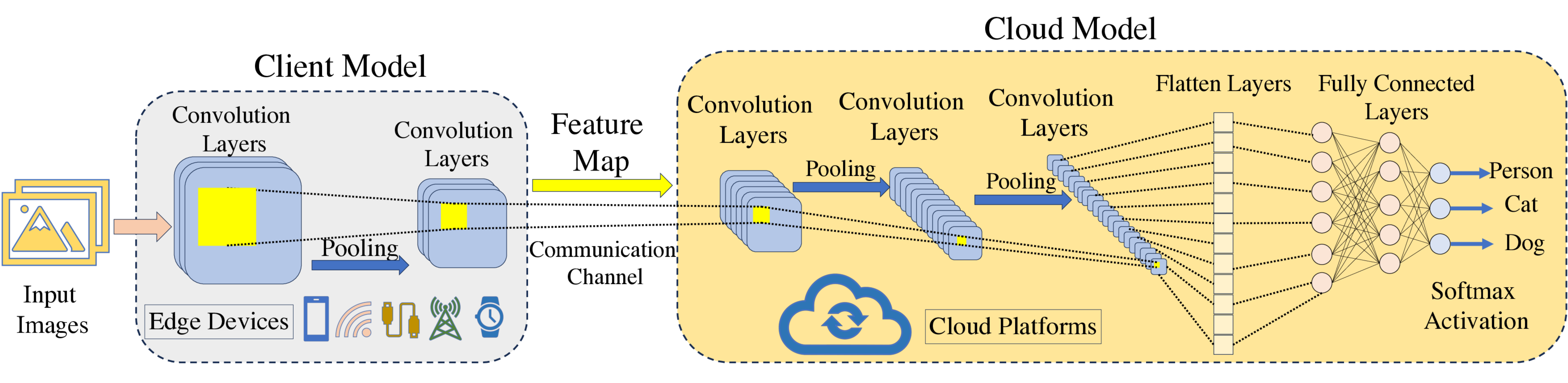}
    \caption{The typical architecture of the $\mathcal{EC}$ system.} 
    \label{fig:arc}
\end{figure*}

In recent years, a new computational architecture, $\mathcal{EC}$ system, has received considerable attention in both academia and industries. Rather than deploying the whole model on the edge device, in the $\mathcal{EC}$ system, a deep neural network (DNN) is split into two components. The shallow part of the DNN model is deployed on the edge devices, whereas the larger portion is installed on the cloud side \citep{wang2018not}. In $\mathcal{EC}$ systems, clients' raw data is fed into the client-side model and transformed into abstract feature maps. Then, these feature maps are sent to the cloud-side model for the final inference and prediction \citep{he2020attacking, wang2022pcnncec}. Fig. \ref{fig:arc} shows a typical Convolutional Neural Network (CNN) architecture in the $\mathcal{EC}$ systems, where the first two convolutional layers are in the edge devices as the client model, and the rest of the layers (including convolutional layers, flatten layers, and the fully connected layers) are deployed on the cloud side as the cloud model. With the collaboration between edge devices and cloud servers, users cannot only enjoy advanced high-performance AI services without cumbersome computational overheads but also avoid sharing their original local data with other parties. 

Nevertheless, the privacy issues are still unresolved, which hinders the further adoption of the $\mathcal{EC}$ systems into more practical applications. To be specific, the feature maps transferred between the edge devices and the cloud platforms still contain sensitive information. By taking advantage of advanced attacking methods, adversaries and malicious cloud servers can infer private knowledge from the transferred feature maps. For instance, model inversion attacks can be performed in the $\mathcal{EC}$ system to reconstruct images in an image classification system \citep{he2020attacking, he2019model, gu2018confidential, yang2022security, duan2022distributed}. By querying the client model and generating an adversarial training dataset, attackers can build effective deep learning (DL) models to get sensitive information as well  \citep{wang2019private, osia2020hybrid, chi2018privacy, osia2017privacy, erdougan2022unsplit}.

There are a number of endeavors that aim to protect client's data privacy from being disclosed in the $\mathcal{EC}$ systems, which can be mainly divided into two categories. The first category is called the plug-in-based approaches, which mainly includes the dimensionality reduction (DR) approach and noisy injection. The second category is called the non-plug-in approaches, where adversarial training and model compression are the typical examples. Currently, adversarial training is the most commonly studied solution \citep{jingtao2022ressfl, thapa2022splitfed}. In such approaches, the loss functions are usually modified based on privacy purposes. Essentially, by adding new items in the loss function, the modified network architecture can maintain the primary task as accurately as possible, but reduce the adversarial model's sensitive task performance to a large extent. Here the primary or sensitive task can be classification, regression, or any type of machine learning (ML) task. Typical representations of such techniques are \cite{wang2018not, li2021deepobfuscator, ding2020privacy}. One major drawback of the non-plug-in approaches such as adversarial training is that extra modifications should be performed on the neural network models, which makes the new $\mathcal{EC}$ pipeline not compatible with the existing AI algorithms. Moreover, an adversarial training dataset should be collected and extra training should be conducted before the new models are launched. Compared with the non-plug-in approaches, plug-in techniques are well known for their lightweight and easy-to-perform properties. Considering the add-on component, plug-in solutions usually add random noise into the feature maps, which does not require any data collection or training on the existing DL models. Except for other good characteristics, such as low communication overheads and low computational costs, its limitations are also obvious.  After adding noise, the perturbed feature maps can affect the adversary's attacking model and the cloud-side model at the same time. In other words, although clients' privacy can be preserved, the performance of the primary task will also be degraded and the utility cannot be maintained. 

With many works focusing on non-plug-in privacy-preserving solutions, we found that the plug-in approaches have not been adequately studied in the past. Most of the plug-in methods are used as the baseline techniques for comparison purposes \citep{osia2020hybrid}. In this paper, we propose a novel plug-in-based solution, called Autoencoder-based Delta Protection (ADP), by combining the autoencoder (AE) \citep{zhai2018autoencoder} and Class Activation Maps (CAMs) \citep{zhou2016learning} for protecting the sensitive information in the $\mathcal{EC}$ image inference system. In particular, in attacking simulations, CAMs are first used to find the most essential and important regions in the images where the adversarial model is focusing. Next, a blurring technique is used on the image to filter out the most significant region for the sensitive attribute inference, but preserve the information for the primary task. After that, AE is trained and optimized using the feature maps that are coming from the blurred images and will be for the privacy protection in the $\mathcal{EC}$ platforms. The major contributions of this paper can be specified as follows:

\begin{itemize}
    \item The end-to-end pipeline in the $\mathcal{EC}$ system is implemented, which includes the primary image classification with edge-cloud collaboration architecture, the image reconstruction models based on white-box attack, and the sensitive attribute inference attack. The basic components can support comprehensive model evaluations and results comparison.
    \item An effective plug-in approach is developed and designed for privacy protection in the $\mathcal{EC}$ system. This novel technique is the combination of AE and CAMs, which can provide explainable and attribute-specific privacy protection with all the plug-in properties, without having to retrain or modify the original $\mathcal{EC}$ model. Training of the plug-in protection is allowed. 
    \item A comprehensive set of experiments are conducted to validate the performance of the proposed solutions. The superiority of the proposed novel method with respect to the primary task's performance, image reconstruction, and sensitive attribute inference. Under certain experimental settings, upon applying the proposed protection technology, performance loss at the cloud-side model is around 4\%, whereas the performance drop at the adversarial side is more than 20\%. 
\end{itemize}

The experimental evaluations are based on the state-of-the-art object detection models. The remainder of this paper is organized as follows. Section II reviews related state-of-the-art works. Section III describes the problem statement. Section IV describes our proposed plug-in-based protection approach, which is based on AE and CAMs. This section also explains the Principal Component Analysis (PCA) based protection approach, which is used as a competitor to evaluate the performance of the proposed novel technique. The set-up description, parameter settings, and experimental results are discussed in Section V. Finally, the paper is concluded in Section VI.

\section{Literature Review}

In this section, we review the state-of-the-art privacy-protection solutions in $\mathcal{EC}$ settings. Additionally, we examine similar solutions in non-$\mathcal{EC}$ settings where relevant \citep{ZHANG2025112965}.

Principal Component Analysis (PCA), as a representative dimensionality reduction algorithm, is widely used in several works to protect client-side outputs (i.e., feature maps) \citep{malekzadeh2017replacement, osia2020hybrid, osia2017privacy}.  \cite{osia2020hybrid}, employ PCA as a feature extractor to retain essential information while removing unnecessary details from client-side input (e.g., images). A dense reduction matrix is applied in the final layer to maintain consistent input/output configuration for DL models. 

Another common plug-in approach is the use of noisy injections, employed in various works such as \cite{he2020attacking, mireshghallah2019shredder, mao2020privacy, wang2018not}. For example, in \cite{wang2018not}, Differential Privacy (DP) noise is introduced to transform client-side outputs securely. Specifically, the client-side input is multiplied by a mask matrix for nullification, followed by the addition of differential privacy noise at a particular layer of the client-side model to further enhance privacy. In some cases, noisy injections are combined with adversarial training techniques to increase the model's robustness to random perturbation after training. 

In non-$\mathcal{EC}$ settings, \cite{zheng2023gone} introduce GONE, which uses DP to protect the output class probabilities targeted by adversaries in membership inference and model-stealing attacks. However, this method only protects server-side outputs. Regarding the adversarial training solutions, \cite{li2021deepobfuscator} train four components simultaneously, an \textit{obfuscator}, \textit{classifier}, \textit{adversary reconstructor}, and \textit{adversary classifier}. Similarly, \cite{ding2020privacy} address both the primary image classification task and sensitive attribute inference attacks concurrently. By designing specific loss functions, their proposed framework ensures the primary task minimizes loss while maximizing the loss for attackers. 

\cite{wang2018not} propose a noisy training technique combining differential privacy, generative adversarial training, and transfer learning. During the noisy training phase, both the original and perturbed client-side inputs are sent to the DL model for parameter optimization, making the model more robust and stable in real-world applications. 

Inspired by the Mix-up architecture, \cite{liu2020datamix} introduce a privacy protection technique for the $\mathcal{EC}$ systems, called Datamix. This approach applies mixing and de-mixing operations to both the client-side inputs and the server-side outputs. A key difference of this work is that the entire CNN model is divided into three components instead of two. The client runs both the first and third models as a feature extractor and predictor, respectively. This designed enables the $\mathcal{EC}$ platform to achieve high accuracy, efficiency, and privacy preservation. 

In non-$\mathcal{EC}$ settings, \textit{generative adversarial networks} (GANs) are used by \cite{kairouz2022fur, avidan2022decouple, hung2020autogan, li2020tiprdc, mohammad2024adjustable} to establish an adversarial interaction between an encoder, adversary, and model. In this setup, the encoder learns to apply protection in the form of noise, optimized through a discriminator and generator during training to maximize the utility for the model and minimize the utility for the adversary (i.e., enhancing privacy). The noise applied may serve to mask, distort, censor, obfuscate, decorrelate, decouple, or otherwise separate sensitive features from primary features in the original input, which is presumably drawn from the same training distribution. In some settings the output may not retain the same dimensionality as the input, which can limit its applicability. Additionally, a threshold may be applied during the learning process to control the utility privacy trade-off. The protection mechanism can be either general or specific, depending on whether the adversarial task is known in addition to the strength of the adversary.

Importantly, the deployment of GAN-based protection in non-$\mathcal{EC}$ settings is typically aimed at safeguarding client-side inputs. However, this type of protection is less effective in an $\mathcal{EC}$ system, where client-side inputs, and their respective intermediate representations, remain entirely private. As a result, only client-side outputs require protection, making the proposed protections overly eager. Moreover, the client-side utility is further impacted if a \textit{retraining restriction} is imposed on the client, preventing adjustments for protection-related perturbations in the input. This is distinct from training the protection method itself, which is permissible as long as it is plug-in in nature. Nevertheless, it may be possible to adapt some of the aforementioned GAN-based approaches to protect client-side outputs in an $\mathcal{EC}$ system. To the best of our knowledge, this remains unexplored.

Apart from noisy injection and adversarial training cryptographic techniques \citep{rouhani2018redcrypt, mishra2020delphi, gu2018securing} and model compression \citep{alwani2022decore, lee2020keystone, yang2022cnnpc, zhang2022communication} are also used. As a typical example of cryptographic solutions, \cite{rouhani2018redcrypt} propose a reconfigurable hardware-accelerated framework for empowering a privacy-preserving inference system, called ReDCrypt. In the proposed system, the client can dynamically analyze their data over time without the requirement of queuing the samples to meet a certain batch size. They require neither re-training the AI models nor relying on two non-colluding servers in their system design. By executing the popular Yao's Garbled Circuit protocol, the inference system is divided into two phases, namely, the privacy-insensitive computation at the client-side, and the privacy-sensitive computation between the client and the cloud. Experimental results show that the designed solution can output 57-fold higher throughput per core with no obvious accuracy loss compared with the state-of-the-art solutions. \cite{liu2017oblivious} study a novel solution that can transform an existing DNN model into an oblivious neural network supporting privacy protection for edge-cloud collaborative inference systems. The proposed model outperforms other solutions in terms of response latency and data size. \cite{gu2018securing} aims to protect the security of input data in the edge-cloud collaborative systems based on a trusted execution environment. Potential information exposure in the edge-cloud-based framework is investigated systematically and a secure enclave technique is designed against reconstruction privacy attack. In addition, the adversaries' attacking capacities are quantified with different threat models and attacking strategies.

In summary, each solution has its own advantages and drawbacks. Cryptographic solutions and adversarial training can guarantee the performance of the primary task while protecting the client privacy, but they introduce significant computational and communication overhead. On the other hand, non-$\mathcal{EC}$ solutions, when applied in the $\mathcal{EC}$ setting, are either too lenient, only protecting the server outputs (and thus offering no privacy benefit), or overly aggressive, protecting the client inputs (which reduces utility without providing additional privacy). Model compression is straightforward to deploy on the client model but suffers from low performance, as it is very challenging for a compact model to achieve high classification accuracy. In contrast to prior approaches, we focus plug-in solutions to protect client-side outputs. Without requiring any retraining of the existing CNN models or modifications to the network architecture, our proposed model achieves a reasonable balance between utility and privacy.

\section{Problem Statement}
Split learning can be leveraged to improve the computational performance (or load) by splitting an existing pre-trained model into two halves. The first is placed on the client and the second is placed on the server. A shallower client-side split is preferable since the client is assumed to be a resource-constrained device, Here, an image is first passed through the client-side layers until the given split position. Then the output, a feature map, is passed from the client to the server. The server then passes the feature map through the rest of its server-side layers before reaching the final inference layer. This technique improves the utility on the server side since the server does not have to pass the feature map through the entire architecture. Thus, the server's utility is improved. Most importantly, unless either the client or server weights are retrained or modified the output of the split model at any given split is equivalent to the original non-split model. 

\begin{figure*}
    \centering 
    \includegraphics[width=0.75\textwidth]{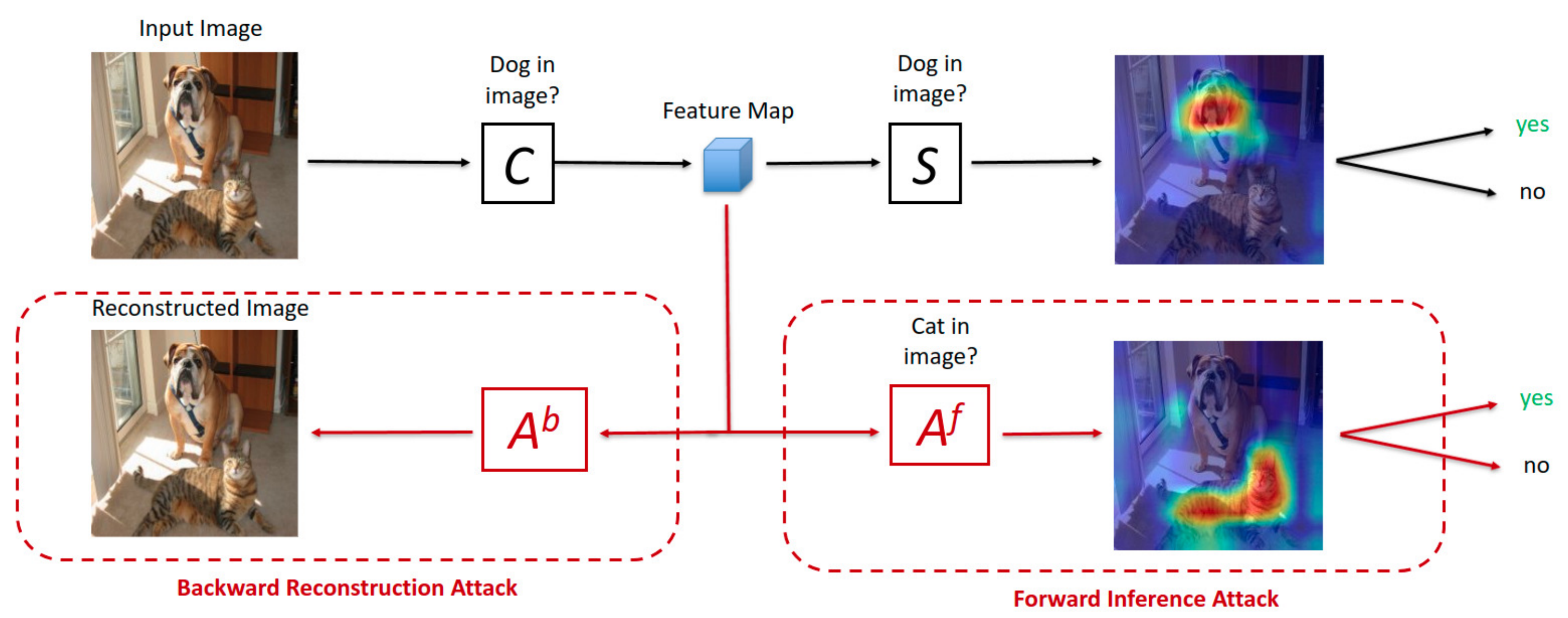}
    \caption{Split model primary and sensitive tasks, partially adapted from \cite{jacobgilpytorchcam}.} 
    \label{fig1}
\end{figure*} 

However, both split and non-split scenarios elicit major privacy concerns as they relate to the information passed from the client to the server. Formally the privacy problem reduces to: \textit{how can the server be trusted to only infer tasks that it primarily purports to instead of other more sensitive tasks?} For example, consider a server model that can accurately detect dogs in images. Also consider a client image sent to the server containing both a cat and a dog in the same image. A curious or malicious server, referred to hereafter as the adversary, could trivially detect if there is a cat in the image by simply using a different model. One that is trained to detect cats instead of dogs. This same scenario is present in the split case but requires that the adversary have the ability to query a shadow client model. The adversary would then uses the shadow model to generate a new labeled dataset of feature maps to labels based on the output of the client $C$. This dataset can then be used to train a server-side adversary model, $A^{f}$, to predict a sensitive task other than that which it purports to. This is defined hereafter as \textit{forward inference attack}. Additionally, an adversary, $A^{b}$, could also attempt to reconstruct the original image. This can be done by passing the received feature map through an inverted client model or another method, defined hereafter as \textit{backward reconstruction attack}. Both these adversarial scenarios, shown in Fig. \ref{fig1}, are realistic since the cloud already acts as a semi-trusted entity with the special knowledge, data, and resources needed to realize the attack that a third party presumably would not. For example, decrypted client-side feature maps, initial training dataset, shadow client model weights, etc. Additionally, in this setting the cloud is recognized to have relatively unbounded computation power to realize the aforementioned attacks, adding to the seriousness of the privacy risks.

To address these privacy risks, it is important to develop protection specifically designed to provide a high degree of privacy while also enabling a high degree of utility. Based on the aforementioned statement, the problem can be formulated in the following manner:
\begin{equation}
\begin{aligned}
\left\{
\begin{aligned}
&S_{\alpha} \approx S_{\beta} \\
&A_{\alpha} \gg A_{\beta} \\
&\lvert{S_{\alpha}} - S_{\beta}\rvert \ll \lvert{A_{\alpha}} - A_{\beta}\rvert \\
\end{aligned}
\right.
\end{aligned}
\label{eq1}
\end{equation}

where $S$ and $A$ stand for the server and adversary accuracy, respectively, and $\alpha$ and $\beta$ indicate before and after protection placed at a given split position. 

However, several challenges, limitations, and trade-offs must be addressed. Firstly, the protection must not involve retraining either the client or server-side model weights as this would violate the previously mentioned equivalency with the original non-split model, presumably leading to unpredictable accuracy or large retraining costs. This means that any solution must be plug-in. Secondly, the plug-in protection must either be placed before the client (front-end protection) acting on the input image or after the client (plug-in protection) acting on the feature map before it is sent to the server. Placing the protection entirely on the server is not feasible as the server cannot be trusted to protect the feature map it receives. Thirdly, due to client and server-side retraining restrictions a strong adversary will always be able to theoretically retrain on the protected feature maps, with any state-of-the-art deep learning architecture, and, thus, have a distinct advantage over the server. Fourthly, a reasonable assessment of the Utility versus Privacy trade-off (server versus adversary accuracy after protection) must occur on a case-by-case basis and consider (1) primary and sensitive task selection, (2) client, server, adversary, and/or protection hyperparameters and architectures, and (3) adversary capabilities.

\section{Plug-in-based Protection Approaches}
When considering solutions to the aforementioned reverse reconstruction and forward inference attacks, a generic protection solution can involve a multitude of statistical, deep learning, and other approaches such as PCA and AE \citep{osia2020hybrid, osia2017privacy}. In the case of deep learning, the protection model must be iteratively trained to convert an input, be it an image or a feature map, into a protected output of matching dimensions. However, the dimensions can vary as the intermediate representations pass through the various hidden layers in the middle of the protection architecture. Additionally, specific data must be carefully selected to achieve high accuracy in the training phase to ensure the protection model generates the correct type of protection. In the inference phase, the protection model is plugged in at the specific split position it was trained on a priori. Afterward, inputs passed to the protection are actively protected as they pass through the various layers on their way to the server model. It is presumed that the input to the protection model is drawn from the same sample context as the original data used to train the protection model. 

\subsection{PCA-based Protection Baseline}
In this baseline plug-in approach, a dimensionality reduction algorithm is used to decompose the feature map onto principal components that capture the maximum variability of the data. In the real-time deployment of the plug-in approach, feature maps of new images that are coming from the client will be reconstructed by means of the extracted principal components in order to minimize the sensitive or private attributes of the feature maps. Here, the reduced Singular Value Decomposition (SVD) algorithm is implemented to perform dimensionality reduction on flattened feature maps from a ResNet50 model split at various bottleneck layers (Bonk). The dimensions of these feature maps are 512 by 28 by 28 (flattened to 401,408). %\cite{1102314,Golub1971} \cite{svd} 

\subsection{Proposed Autoencoder-based Delta Protection (ADP)}
In this section we discuss our proposed plug-in approach that uses both AE and CAMs to protect feature maps with an novel delta $\Delta$ protection.
 
\subsubsection{Autoencoder}
Generally speaking a convolutional AE works similarly to the convolutional layers in both VGG16 and ResNet50 and may contain the same types of supporting layers depending on construction, i.e., Rectified Linear Activation Function (ReLU) layers. One main difference specific to AEs are the use of convolutional transposition or de-convolutional layers that aim to invert a prior convolution operation. However, the inverted result may not be perfectly equivalent, depending on implementation. Importantly, the output dimensions should be the same as the original input to the prior convolutional layer. Generally, AEs use the input and convert it into either a lower or higher dimensional space, and learn the chosen output. Oftentimes the chosen output is the original input. In this way AEs learn to reduce noise in a given input image or complex data to better standardize the inputs such that later inference is easier in the fully connected layers. However, it is possible to instead map the smaller reduced space to a much different output as in our case.

\subsubsection{Class Activation Maps}
CAMs allow for the visual inspection of what a deep learning model uses to infer a given class-based prediction of a given input image. Here, heatmaps relating to the last or multiple convolutional layers of a deep learning model are used to discern what the most important parts of an image are in the final prediction by projecting \textit{hot} and \textit{cold} areas over the input image. Generally, CAMs are used to debug models and explain how deep-learning models make predictions after their weights are fully or partially trained. In our work, we adapt code from \cite{jacobgilpytorchcam} for CAM generation.

\subsubsection{The Proposed Delta Approach}
\begin{figure*}[h]
	\centering 
	\includegraphics[width=0.75 \textwidth]{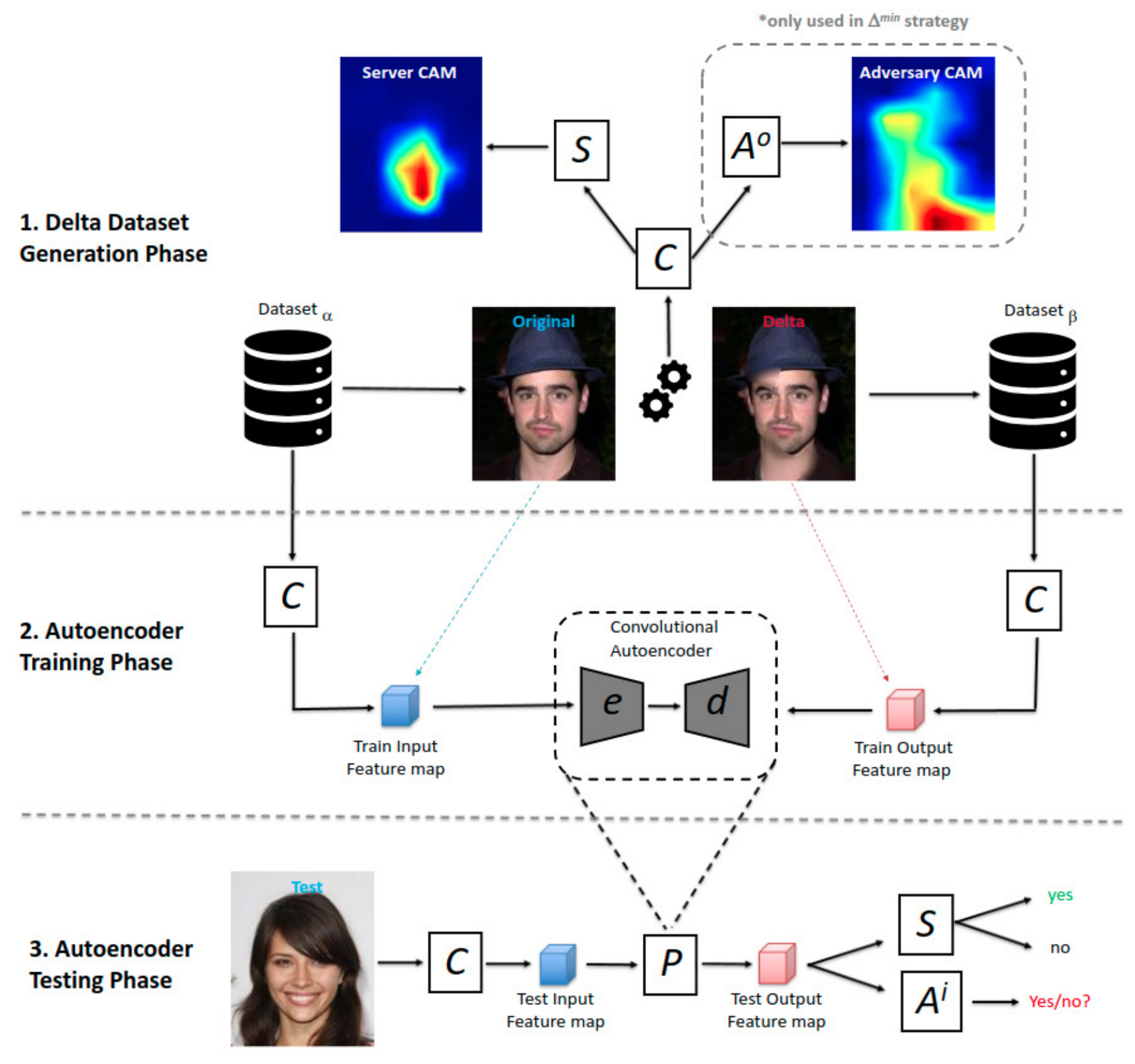}
	\caption{The general diagram of the proposed novel delta approach, called ADP.} 
	\label{fig:delta_strategy}
\end{figure*}

In our novel delta $\Delta$ approach, shown at a high level in Figure \ref{fig:delta_strategy}, the use of CAMs is combined with AEs so that a protection, $P$, mapping between original and protected feature maps can be learned, and, then, applied in real time. Here, two delta strategies are considered, the first where the sensitive adversary tasks are known or can be inferred, and the second where they cannot. In the former strategy, CAMs of both the offline adversary, $A^{o}$, and the server are first used to construct the protected image through a process we generally refer to as $\Delta^{min}$, shown in Algorithm \ref{alg:one}.

\RestyleAlgo{ruled}
\LinesNumbered
% \SetKwComment{Comment}{#}{}
\SetKwFor{RepTimes}{repeat}{times}{end}

\begin{algorithm}[h]
\caption{Delta $\Delta^{min}$ Strategy}\label{alg:one}
\KwData{$n \ge 1$, Input image $I$, Client-model $C$, \\Server-model $S$, Offline-adversary $A_{o}$, Model concatenation operator $\oplus$, Class Activation Map function CAM, CAM negation function $\neg$, Protection intersection function $\Delta$, Temperature threshold $t$, Intensity threshold $i$, Protection method $\delta$ (either ``black-out" or ``blur-out''), Original and protected client-side feature map $F_{o}$ and $F_{p}$}
\KwResult{$F_{o}, F_{p}$}
$F_{o} \gets C(I)$\;
\RepTimes{$n$}{
  $M_{A_o} \gets \text{CAM}(I, C \oplus A_{o})$\;
  $M_{S} \gets \neg\ \text{CAM}(I, C \oplus S)$\;
$I \gets \Delta^{min}(I, M_S, M_{A_o}, t, i, \delta)$\;
}
$F_{p} \gets C(I)$\;
\end{algorithm}

This process involves protecting the intersection of the \textit{cold-server} and \textit{hot-adversary} CAM spaces according to a predefined temperature threshold. Afterward, the protected space is projected back over the original input image to create the final protected image. Both images are then passed thought the client where the original feature map is passed as input to the encoder $e$ and the protected feature map is passed to the decoder $d$ such that the protection can be learned. This strategy can be reapplied in a $\Delta_n^{min}$ scheme in which $n$ stands for number of iterations, where the protected image of the previous delta simply becomes the input to the next. Here, each successive delta is thought to apply more protection since previously targeted hot-adversary regions in the CAM are protected, which generally forces the adversary CAM to change location in contrast to the server regions that do not. This is of course by design as the hot-adversary regions are specifically targeted for protection while the hot-server regions are purposely left unprotected. However, it is entirely possible that the hot-adversary and hot-server regions may intersect. If this happens, we simply prioritize utility over privacy and opt not to protect the intersecting region as doing so will likely cause a significant server-side accuracy drop. Therefore, the feasibility of providing protection with this, \textbf{or any}, delta strategy must consider the degree to which primary and sensitive attribute hot CAM regions intersect. In the second strategy, where the adversary tasks are unknown, protection is only applied to the cold-server CAM regions also predefined by a given temperature threshold, shown in Algorithm \ref{alg:two}. We define this strategy as $\Delta^{max}$ since it protects the maximal amount of information not directly used by the server according to the server CAM. Here, a $\Delta_{n}^{max}$ scheme could theoretically be applied but is not explored in our work.

\begin{algorithm}[h]
\caption{Delta $\Delta^{max}$ Strategy}\label{alg:two}
\KwData{$n \ge 1$, Input image $I$, Client-model $C$, \\Server-model $S$, Model concatenation operator $\oplus$, Class Activation Map function CAM, CAM negation function $\neg$, Protection intersection function $\Delta$, Temperature threshold $t$, Intensity threshold $i$, Protection method $\delta$ (either ``black-out" or ``blur-out''), Original and protected client-side feature map $F_{o}$ and $F_{p}$}
\KwResult{$F_{o}, F_{p}$}
$F_{o} \gets C(I)$\;
\RepTimes{$n$}{
  $M_{S} \gets \neg\ \text{CAM}(I, C \oplus S)$\;
  $I \gets \Delta^{max}(I, M_S, t, i, \delta)$\;
}
$F_{p} \gets C(I)$\;
\end{algorithm}

Furthermore, the specific protection method used to define and protect regions in a given protection strategy falls into two categories. The first, \textit{black-out}, consists of setting all values in the protected space to the same value, in our case black. Here, a threshold intensity value is used to define the protected space by creating a boolean mapping of hot or cold regions. This is accomplished by measuring the blue value in each input CAM pixel to determine if it is less than or greater than the chosen threshold value. The second method, \textit{blur-out}, uses the same threshold intensity to define the protected space but also applies blur protection according to a given blur intensity. Either protection method may be applied to either of the two protection strategies previously mentioned with a high degree of flexibility depending on protection requirements.

After selecting a given combination of delta method, strategy, and applicable parameters a dataset mapping original to protected images can be created.  However, since AE will not be accepting input images directly unless deployed as a front-end protection, both the original image and protected image need to be converted to feature maps. This is accomplished by passing both the original and protected images through the client separately at a given split position in order to obtain the respective feature maps. AE can then train on the feature maps to theoretically learn the protection mapping, ultimately derived from the CAMs of the client, server, and offline adversary (in the $\Delta^{min}$ case), respectively. In AE training process, several types of architectures can be constructed to either increase or decrease the Z dimension of the feature maps. Additionally, the convolution and de-convolutional layers in AE may additionally shrink or expand the X and Y feature map size. Since feature map dimensions may change depending on layer and architecture it may be necessary to create custom AEs that vary the stride, padding and dilation when required if they need to be especially deep. Other parameters such as learning rate, loss function, optimizer, number of epochs, etc. can also be adjusted accordingly. All that remains after this is to deploy AE at the given split where it will take input feature maps and apply the protection mapping it learned during the training phase in the testing phase.

\subsection{Split Position and Feature Map Considerations}
While different deep learning architectures operate with vast differences in terms of the type, number, and configuration of layers several high-level observations of the architectures are worth noting. Firstly, in the case of VGG16 and ResNet50 both architectures take a color image as input. Here, the image is a three-dimensional tensor representing the red, green, and blue dimension combination values used to display color at every pixel in an image. Importantly, this input image is merely a special case feature map, one that has yet to pass through any given layer. As such, it still has the exact same properties as later feature maps. Namely, a Z or depth dimension, and an X and Y or height and width dimension for every Z dimension. Secondly, both architectures progressively perform a series of operations that increase the depth of the feature map while \textit{generally} reducing the height and width of each depth dimension. The main aim of this general approach is to recursively reduce the overall amount of information while retaining only the most important information in the feature extraction phase. This is needed since in the latter classification phase the fully-connected layers need a reduced feature space. Importantly, feature map slices or X and Y cross sections are more representative of the original input image at shallower layers, and hence split positions, since they will have greater height and width dimensions and increased overall information.

These observations are very important from several implementation and theoretical perspectives. Firstly, from an implementation perspective, any dataset that could be used to train or fit a given adversarial or protection model comprised of feature maps rather than images will grow by a constant factor of the depth of the feature maps generated at a given split position relative to the same dataset of input images. This is especially important since attempting to hold even a moderate dataset of feature maps in memory or on disk in many cases is not feasible. Typically, during the training session, feature maps must be computed from their respective input images by passing them through the client every time they are used. Secondly, from a theoretical perspective, it is much easier to protect feature maps at deeper split positions for a multitude of reasons. Specifically, feature maps at later positions, while greater in number, individually contain less information making both reconstruction and inference attacks increasingly difficult or in extreme cases infeasible beyond acceptable thresholds. While it is tempting to use this observation na\"ively as a direct form of protection it unfortunately contradicts the utility provided by the nature of the split learning to begin with as it places most of the layers on the resource-constrained client. However, the split position still plays a very important role in the efficacy of implemented protections and can significantly increase or reduce the general challenges associated with providing adequate protection.

\section{Experimental Results} \label{sec:results}

Here, the experimental setting is initially explained, and, then, the achieved results through each privacy protection strategy are presented.

\subsection{ADP Experimental Setup}

Initially, the process of data generation, training, and testing are explained, and, then, network architecture and its parameter settings are reported. 

\subsubsection{Dataset Generation}

Dataset generation in our work follows a scheme, where $n$ and $m$ binary samples are evenly selected from the CelebFaces Attributes Dataset (CelebA). Samples are selected according to a given primary or sensitive attribute without replacement to ensure that there are no overlapping train and test samples in the respective datasets. Primary and sensitive attribute overlap is not considered, consistent with the real-world adversary case. The delta dataset follows a similar scheme but is generated separately and aggregates \textit{n} evenly selected binary samples for one or more attributes. Both sensitive or primary attribute samples can be used.

\subsubsection{Protection Training and Testing}

When testing the delta approach with any combination of strategy, method, or other parameters, it is important to consider several factors that impact protection training and testing. Firstly, it is important to recognize that the delta dataset generation process is strongly dependent on the server $S$, adversary $A$, and especially the client model $C$. This is the case since the client model is used for converting original and protected samples to feature maps while all models are used via their CAMs (that are strongly associated with the weights of the various models) to create the protection training samples directly. In the ideal case, a separate set of client, server, and adversary models should be used to train and test AE protection. This approach would theoretically indicate if AE protection is generalizable enough to plug into any given client-server model or if it only affords protection to the specific client-server model it was trained on. Similar generalizability would also extend to adversaries that presumably have the same task and indicate if the protection only applies to the specific adversary the protection was trained on. However, we consider that the protection ought only to apply to the client-server model that the protection was trained on since it has an undue influence over the CAMs used to generate the protection. Consequently, we do consider both training and testing adversaries, referred to hereafter as offline and inference adversaries $A^{o}$ and $A^{i}$, respectively, to evaluate the protection in order to strike an acceptable level of adversarial generalizability. Importantly, the offline adversary model need not be the same as the inference adversary model. Either model may utilize the existing server-side architecture or an entirely different, and presumably deeper, architecture. We refer to these two adversary cases generally as \textit{split} and \textit{full}, respectively. Additionally, in the full case, the arbitrary architecture must be able to accept a feature map of Z dimension after a given client split and have a large enough remaining feature map of X and Y dimensions to pass through the remaining architecture without being reduced to zero by the remaining convolutional layers. Nonetheless, both offline and inference adversaries need to be trained on the same client-server and have similar, if not identical, sensitive tasks. The client-server will always fall under the \textit{split} case by definition.

\subsubsection{Experiment Architecture}

In our work, we employ a pipeline architecture to carry out various types of experiments. This is the case since many aspects of testing are strongly dependent on each other. The client-server training must proceed with adversary training, delta dataset creation must proceed with protection training, and so on. However, there are certain benefits to this architecture as certain sections of the pipeline can be cached and reused in consecutive experiments while others sections can be run in parallel.

\begin{table}[htpb]
\caption{Parameter setting for VGG16 and ResNet50 experiments through autoencoder.}
\centering
\resizebox{\columnwidth}{!}{
\begin{tabular}{lll}
\hline\noalign{\smallskip}
Experimental Parameters & VGG16 & ResNet50 \\
\hline\noalign{\smallskip}
Primary Attribute & Wearing\_Lipstick & Wearing\_Lipstick \\
Primary Train Dataset & 2000 balanced samples & 1000 balanced samples\\
Primary Test Dataset & 400 balanced samples & 200 balanced samples\\
Sensitive Attribute & Wearing\_Hat & Wearing\_Hat\\
Sensitive Train Dataset & 2000 balanced samples & 1000 balanced samples\\
Sensitive Test Dataset & 400 balanced samples & 200 balanced samples\\
Delta Dataset  & 4000  & 2000\\ % balanced samples with both sensitive and primary attributes
Delta Threshold Intensity & 99\% & 99\%\\
Delta Blur Intensity & 40 & 40\\
$\Delta^{min}_{n}$ & 2 & 2\\
Architecture & VGG16 & ResNet50\\
Split Positions & Convolutional layers \{4,8,12\} & Bottleneck layers \{4,8,12\}\\
Reconstruction Iterations & 20,000 & ---\\
\hline\noalign{\smallskip}
\end{tabular}
}
\label{tab:settings}
\end{table}

Here, we provide the setup parameters, shown in Table \ref{tab:settings}, as they relate to the experiments conducted on the VGG16 and ResNet50 architectures using the ADP approach. The main difference between the VGG16 and ResNet50 parameters is that the VGG16 parameters use bigger datasets to achieve higher accuracy as the VGG16 architecture is smaller compared to the ResNet50 model. 

\subsubsection{Autoencoder Architecture}

In our work, we utilize several AEs that decrease the size of the feature map depending on if a decreasing, decreasing\_deep, or decreasing extra\_deep AE is chosen. In the case of the decreasing AE, only one layer is used to reduce or encode the Z dimension of the feature map by a factor of two before it is decoded to the prior dimensions. With the increasing depth of the AE, the reversibility becomes an issue depending on the input dimensions of the feature map. As such not all AEs work on all layers of all architectures. This is the case for Table \ref{tab:aeresinf_deep} that only includes bottleneck layer 4. However, it is possible to create specific AEs for specific layers at the cost of comparability when evaluating results, though we do not do so here. In the case of the decreasing\_deep and decreasing\_extra\_deep AE, they reduce the Z dimension of the feature maps by a factor of four and eight, respectively. In our work, we apply the decreasing AEs to all architectures and select layers as we only apply the decreasing\_deep and decreasing\_extra\_deep AE to the shallowest layer, where they are able to revert the feature map back to the correct output dimensions for the tested split positions.

\subsection{PCA Experimental Setup}

In this experiment, the PCA-based plug-in strategy is evaluated with the experimental setup, as reported in Table \ref{tab:pcasettings}, to match the AE experimental setup. 

\begin{table}[htpb]
\caption{Parameter setting for the ResNet50 experiment through PCA.}
\centering
\begin{tabular}{ll}
\hline\noalign{\smallskip}
Experimental Parameters & PCA \\
\hline\noalign{\smallskip}
Primary Attribute & Wearing\_Lipstick \\
Sensitive Attribute & Wearing\_Hat \\
Train Dataset & 1000 balanced samples \\
Test Dataset  & 200 balanced samples \\
Adversary Test Dataset & 200 balanced samples \\
Adversary Architecture & ResNet50 split \\
Number of Components & 1000 \\
\hline\noalign{\smallskip}
\end{tabular}
\label{tab:pcasettings}
\end{table}

All components used for reconstruction are selected from the start (i.e., from index 0).

\subsection{Evaluation Criteria}

In this work, the effectiveness of various attacks and proposed protections is evaluated using several criteria. In the forward inference attack, the server and the adversary accuracy before and after protection are used to determine if Equation \ref{eq1} is satisfied. For the backwards reconstruction attack the Multi-scale Structured Similarity Index (MS-SSIM) \citep{msssim} is used, which is an extension of the Structural Similarity Index (SSIM) \citep{ssim}, that given two images, in our case original $\phi$ and reconstructed $\theta$, it provides a good automatic approximation of how well a human could distinguish the difference in quality between the two as follows:

\begin{equation}
\text{MS-SSIM}_{(\phi,\theta)} = [l_{M}(\phi,\theta)]^{\zeta_{M}}{\prod_{j=1}^{M}}[c_j(\phi,\theta)]^{\eta_j}[s_j(\phi,\theta)]^{\gamma_j} \\
\end{equation}

\begin{equation}
l(\phi, \theta) = \frac{2\mu_{\phi}\mu_{\theta}+\Gamma_{1}}{\mu_{\phi}^2+\mu_{\theta}^2+\Gamma_{1}}, \\
\end{equation}

\begin{equation}
c(\phi, \theta) = \frac{2\sigma_{\phi}\sigma_{\theta}+\Gamma_{2}}{\sigma_{\phi}^2+\sigma_{\theta}^2+\Gamma_{2}}, \\
\end{equation}

\begin{equation}
s(\phi, \theta) = \frac{\sigma_{\phi\theta}+\Gamma_{3}}{\sigma_{\phi}\sigma_{\theta}+\Gamma_{3}}, \\
\end{equation}

\begin{equation}
\Gamma_{1} = (\Theta_1\Xi)^2, \Gamma_{2} = (\Theta_2\Xi)^2, \Gamma_{3} = \Gamma_{2}/2 \\
\end{equation}

MS-SSIM computes this similarity by considering luminance $l$, contrast $c$, and structure $s$ at various scales $M$ and combing them using a normalized equation. Additionally, $\mu_{\phi}$, $\mu_{\theta}$, $\sigma_{\phi}$, and $\sigma_{\theta}$, are the mean and variance of $\phi$ and $\theta$, respectively, (indicating luminance and contrast) in addition to $\sigma_{or}$ covariance (indicating structure). $\Gamma_{1}$, $\Gamma_{2}$, and $\Gamma_{3}$ are used as small constants, where $\Xi$ is the dynamic range of pixel values and $\Theta_1$ and $\Theta_2$ are small scaler values less than one. The exponents $\zeta$, $\eta$, and $\gamma$, are used to adjust the relative importance of luminance, contrast, and structure components respectively. 

\subsection{ADP Results}

In the forward inference attack, experiment values are listed in Tables~\ref{tab:aevgginf}-\ref{tab:aeresinf_deep} show the attained accuracies for the server before and after protection, $S_\alpha$ and $S_\beta$, as well as offline, $A^{o}_\alpha$ and $A^{o}_\beta$, and inference $A^{i}_\alpha$ and $A^{i}_\beta$ adversaries before and after protection. Various architectures, split positions, and protections are also taken into account. Additionally, each unique combination of protection architecture (decrease and decrease\_extra\_deep) and delta method (Blur and Black) are reported as an independent experiment. However, the corresponding $\Delta^{min}$ and $\Delta^{max}$ strategies are dependently tested on the same client-server architecture and datasets. Some level of accuracy fluctuation across independent experiments is expected through random chance alone but are comparable as they are drawn from the same dataset distribution according to the same random process. Also, a special case exists where $A^{o}_\alpha$ and $A^{o}_\beta$ values are reported in the $\Delta^{max}$ case even though they have no influence on AE training. Here, we still list these results as they can give insight into testing a split or full adversary architecture or simply validate the results. 

\begin{figure*}[htpb]
  \centering
  \begin{subfigure}{0.3\textwidth}
      \includegraphics[width=\textwidth]{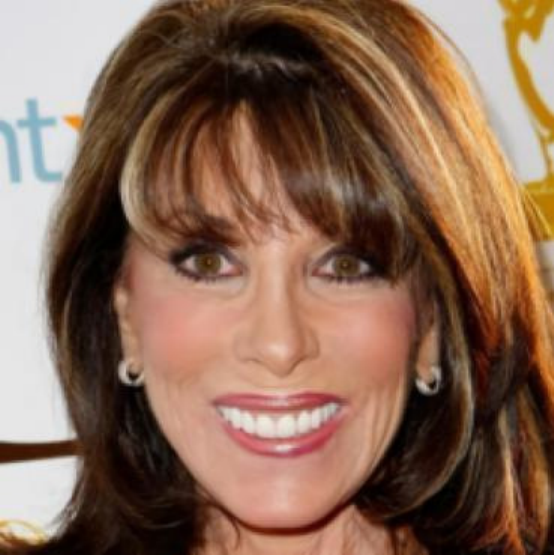}
      \caption{}
      \label{fig:first}
  \end{subfigure}
  \hspace{0.005\textwidth}
  \begin{subfigure}{0.3\textwidth}
      \includegraphics[width=\textwidth]{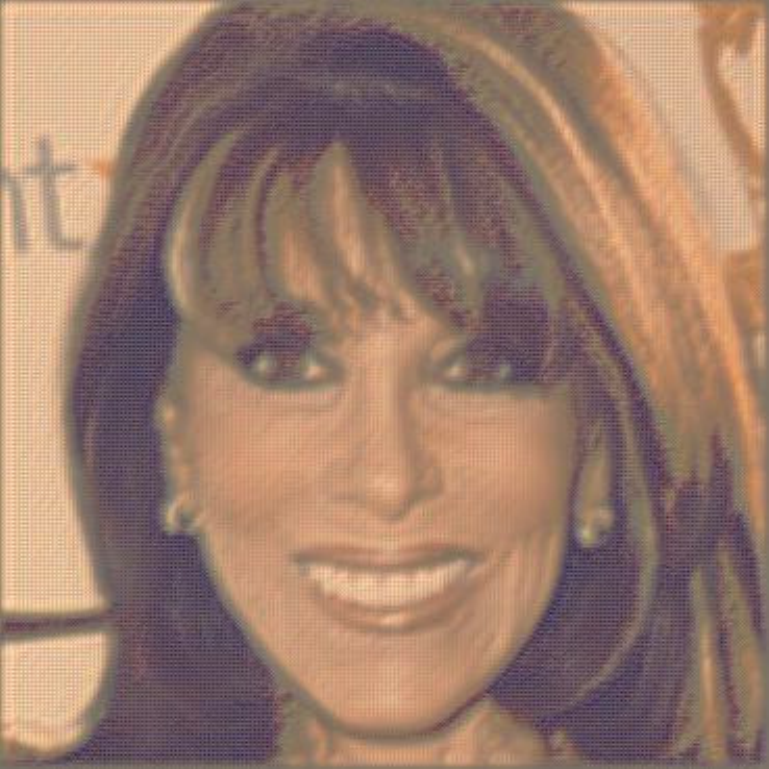}
      \caption{}
  \end{subfigure}
  \hspace{0.005\textwidth}
  \begin{subfigure}{0.3\textwidth}
      \includegraphics[width=\textwidth]{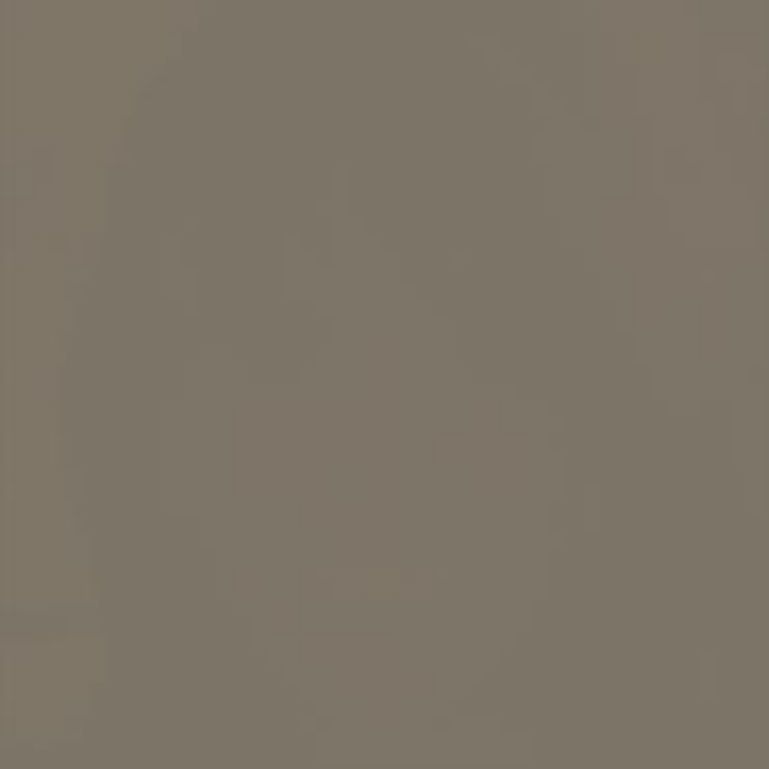}
      \caption{}
  \end{subfigure}
  \caption{Example of backward reconstruction attack on VGG16 convolutional layer 4, with original image shown in (a), reconstructed image in (b), and protected image in (c). The protection is provided using the ADP strategy: $\Delta_{2}^{min}$ and delta method: blur-out along with an offline adversary architecture: VGG16 split and inference adversary architecture: VGG16 split.}
  \label{fig:overall}
\end{figure*}

In the backward reconstruction attack we list MS-SSIM experiment values in Table \ref{tab:aevggrec} before and after protection. Additionally, various split positions and protections as those in the forward inference attack case are listed. We only consider the VGG16 architecture in the backward reconstruction attack case since it is the only architecture we can use to attack with successful results without protection. Lastly, a single preselected image, shown in Figure \ref{fig:overall}, is used to measure the reconstruction threshold before and after protection in all reported experiments. This is the case as the backward reconstruction attack only considers one image at a time and is computationally intensive.

\begin{table}[htpb]
\caption{VGG16 backward reconstruction attack results for primary task: Wearing\_Lipstick and sensitive task: Wearing\_Hat with 400 test samples, 2000 train samples, and 2000 delta samples. $C$, $S$, and $A^{o,i}$ are trained for 250 epochs. All models utilize the corresponding \textit{split} architecture. Autoencoder protection is trained for 120 epochs with a \textbf{decreasing} architecture.}
\centering
\resizebox{\columnwidth}{!}{
\begin{tabular}{ccccc}
\hline
$\Delta$   & $\Delta$ & \multirow{2}{*}{$Split$} & MS-SSIM            & MS-SSIM\\
$Strategy$ & $Method$ &                          & Without Protection & With Protection \\
\hline
& & Conv04 & 0.80 & 0.18 \\
$\Delta_{2}^{min}$ & Blur-out & Conv08 & 0.63 & 0.23 \\
& & Conv12 & 0.57 & 0.23 \\
\hline
& & Conv04 & 0.75 & 0.18 \\
$\Delta_{2}^{min}$ & Black-out & Conv08 & 0.63 & 0.20 \\
& & Conv12 & 0.57 & 0.20 \\
\hline
& & Conv04 & 0.80 & 0.18 \\
$\Delta^{max}$ & Blur-out & Conv08 & 0.63 & 0.18 \\
& & Conv12 & 0.57 & 0.18 \\
\hline
& & Conv04 & 0.75 & 0.17 \\
$\Delta^{max}$ & Black-out & Conv08 & 0.63 & 0.17 \\
& & Conv12 & 0.57 & 0.16 \\
\hline
\end{tabular}
}
\label{tab:aevggrec}
\end{table}

\begin{table}[htpb]
\caption{VGG16 forward inference attack results using the same experimental parameters as in \ref{tab:aevggrec}, except using a \textbf{decreasing\_extra\_deep} architecture.}
\centering
\resizebox{\columnwidth}{!}{
\begin{tabular}{ccccccccc}
\hline
$\Delta$   & $\Delta$    & \multirow{2}{*}{$Split$} & \multirow{2}{*}{$S_\alpha$} & \multirow{2}{*}{$S_\beta$} & \multirow{2}{*}{$A^{o}_\alpha$} & \multirow{2}{*}{$A^{o}_\beta$} & \multirow{2}{*}{$A^{i}_\alpha$} & \multirow{2}{*}{$A^{i}_\beta$} \\
$Strategy$ & $Method$ \\
\hline
& & Conv04 & 87\% & 86\% & 93\% & 93\% & 90\% & 89\% \\
$\Delta_{2}^{min}$ & Blur-out  & Conv08 & 87\% & 86\% & 92\% & 91\% & 88\% & 87\% \\
& & Conv12 & 87\% & 87\% & 90\% & 90\% & 87\% & 87\% \\
\hline
& & Conv04 & 89\% & 89\% & 93\% & 88\% & 94\% & 83\% \\
$\Delta_{2}^{min}$ & Black-out & Conv08 & 89\% & 80\% & 91\% & 84\% & 91\% & 88\% \\
& & Conv12 & 89\% & 66\% & 93\% & 71\% & 89\% & 73\% \\
\hline
& & Conv04 & 87\% & 79\% & 93\% & 77\% & 90\% & 76\% \\
$\Delta^{max}$ & Blur-out  & Conv08 & 87\% & 72\% & 92\% & 85\% & 88\% & 80\% \\
& & Conv12 & 87\% & 70\% & 90\% & 84\% & 87\% & 80\% \\
\hline
& & Conv04 & 89\% & 50\% & 93\% & 73\% & 94\% & 50\% \\
$\Delta^{max}$ & Black-out & Conv08 & 89\% & 50\% & 91\% & 51\% & 91\% & 50\% \\
& & Conv12 & 89\% & 49\% & 93\% & 50\% & 89\% & 51\% \\
\hline
\end{tabular}
}
\label{tab:aevgginf}
\end{table}

\begin{table}[htpb]
\caption{ResNet50 forward inference attack results for primary task: Wearing\_Lipstick and sensitive task: Wearing\_Hat with 200 test samples, 1000 train samples, and 1000 delta samples. $C$, $S$, and $A^{o,i}$ are trained for 100 epochs. All models utilize the corresponding \textit{split} architecture unless indicated otherwise as \textit{full} in bold. Autoencoder is trained for 60 epochs with a \textbf{decreasing} architecture. Values in blue and red correspond to server and inference adversary values reported in Figure \ref{fig:subplot_layout}(a-c), respectively, at epoch 60.}
\centering
\resizebox{\columnwidth}{!}{
\begin{tabular}{ccccccccc}
\hline
$\Delta$   & $\Delta$    & \multirow{2}{*}{$Split$} & \multirow{2}{*}{$S_\alpha$} & \multirow{2}{*}{$S_\beta$} & \multirow{2}{*}{$A^{o}_\alpha$} & \multirow{2}{*}{$A^{o}_\beta$} & \multirow{2}{*}{$A^{i}_\alpha$} & \multirow{2}{*}{$A^{i}_\beta$} \\
$Strategy$ & $Method$ \\
\hline
& & Bonk04 & 88\% & 82\% & 88\% & 69\% & 90\% & 88\% \\
$\Delta_{2}^{min}$ & Blur-out & Bonk08 & 88\% & 87\% & 88\% & 86\% & 90\% & 89\% \\
& & Bonk12 & 88\% & 87\% & 88\% & 86\% & 90\% & 89\% \\
\hline
& & Bonk04 & 84\% & 82\% & 88\% & 81\% & 87\% & 79\% \\
$\Delta_{2}^{min}$ & Black-out & Bonk08 & 84\% & 81\% & 88\% & 84\% & 88\% & 73\% \\
& & Bonk12 & 84\% & 87\% & 88\% & 79\% & 88\% & 77\% \\
\hline
& & Bonk04 & 77\% & 78\% & \textbf{62\%} & \textbf{62\%} & 75\% & 72\% \\
$\Delta_{2}^{min}$ & Blur-out & Bonk08 & 77\% & 74\% & \textbf{69\%} & \textbf{63\%} & 78\% & 72\% \\
& & Bonk12 & 77\% & 77\% & \textbf{71\%} & \textbf{70\%} & 77\% & 60\% \\
\hline
& & Bonk04 & \textcolor{blue}{84\%} & \textcolor{blue}{75\%} & \textbf{65\%} & \textbf{46\%} & \textcolor{red}{71\%} & \textcolor{red}{53\%} \\
$\Delta_{2}^{min}$ & Black-out & Bonk08 & \textcolor{blue}{84\%} & \textcolor{blue}{85\%} & \textbf{68\%} & \textbf{62\%} & \textcolor{red}{73\%} & \textcolor{red}{50\%} \\
& & Bonk12 & \textcolor{blue}{85\%} & \textcolor{blue}{77\%} & \textbf{63\%} & \textbf{57\%} & \textcolor{red}{72\%} & \textcolor{red}{65\%} \\
\hline
& & Bonk04 & 88\% & 52\% & 88\% & 60\% & 90\% & 75\% \\
$\Delta^{max}$ & Blur-out & Bonk08 & 88\% & 49\% & 88\% & 62\% & 90\% & 68\% \\
& & Bonk12 & 88\% & 58\% & 88\% & 60\% & 90\% & 70\% \\
\hline
& & Bonk04 & 84\% & 48\% & 88\% & 56\% & 87\% & 51\% \\
$\Delta^{max}$ & Black-out & Bonk08 & 84\% & 51\% & 88\% & 68\% & 88\% & 62\% \\
& & Bonk12 & 84\% & 50\% & 88\% & 50\% & 87\% & 56\% \\
\hline
& & Bonk04 & 77\% & 58\% & \textbf{62\%} & \textbf{57\%} & 75\% & 57\% \\
$\Delta^{max}$ & Blur-out & Bonk08 & 77\% & 59\% & \textbf{69\%} & \textbf{63\%} & 78\% & 65\% \\
& & Bonk12 & 77\% & 78\% & \textbf{71\%} & \textbf{59\%} & 77\% & 59\% \\
\hline
& & Bonk04 & 84\% & 49\% & \textbf{65\%} & \textbf{50\%} & 71\% & 50\% \\
$\Delta^{max}$ & Black-out & Bonk08 & 84\% & 49\% & \textbf{68\%} & \textbf{46\%} & 73\% & 50\% \\
& & Bonk12 & 85\% & 49\% & \textbf{63\%} & \textbf{48\%} & 72\% & 50\% \\
\hline
\end{tabular}
}
\label{tab:aeresinf}
\end{table}

\begin{table}[htpb]
\caption{Resnet50 forward inference attack results using the same experimental parameters as in \ref{tab:aeresinf}, except using a \textbf{decreasing\_extra\_deep} architecture.}
\centering
\resizebox{\columnwidth}{!}{
\begin{tabular}{ccccccccc}
\hline
$\Delta$   & $\Delta$    & \multirow{2}{*}{$Split$} & \multirow{2}{*}{$S_\alpha$} & \multirow{2}{*}{$S_\beta$} & \multirow{2}{*}{$A^{o}_\alpha$} & \multirow{2}{*}{$A^{o}_\beta$} & \multirow{2}{*}{$A^{i}_\alpha$} & \multirow{2}{*}{$A^{i}_\beta$} \\
$Strategy$ & $Method$ \\
\hline
$\Delta_{2}^{min}$ & Blur-out & Bonk04 & 81\% & 64\% & 88\% & 61\% & 90\% & 52\% \\
$\Delta_{2}^{min}$ & Black-out & Bonk04 & 79\% & 57\% & 91\% & 49\% & 89\% & 67\% \\
$\Delta_{2}^{min}$ & Blur-out & Bonk04 & 84\% & 53\% & \textbf{61\%} & \textbf{57\%} & 76\% & 57\% \\
$\Delta_{2}^{min}$ & Black-out & Bonk04 & 78\% & 50\% & \textbf{61\%} & \textbf{60\%} & 78\% & 50\% \\
\hline
$\Delta^{max}$ & Blur-out & Bonk04 & 81\% & 55\% & 88\% & 56\% & 90\% & 64\% \\
$\Delta^{max}$ & Black-out & Bonk04 & 79\% & 49\% & 91\% & 76\% & 89\% & 50\% \\
$\Delta^{max}$ & Blur-out & Bonk04 & 84\% & 57\% & \textbf{61\%} & \textbf{55\%} & 76\% & 56\% \\
$\Delta^{max}$ & Black-out & Bonk04 & 78\% & 49\% & \textbf{61\%} & \textbf{50\%} & 78\% & 50\% \\
\hline
\end{tabular}
}
\label{tab:aeresinf_deep}
\end{table}

\begin{figure*}[ht]
  \centering
  \hspace{-1em}
  \begin{subfigure}{0.45\textwidth}
    \centering
    \includegraphics[width=\linewidth, trim=50 10 100 50, clip]{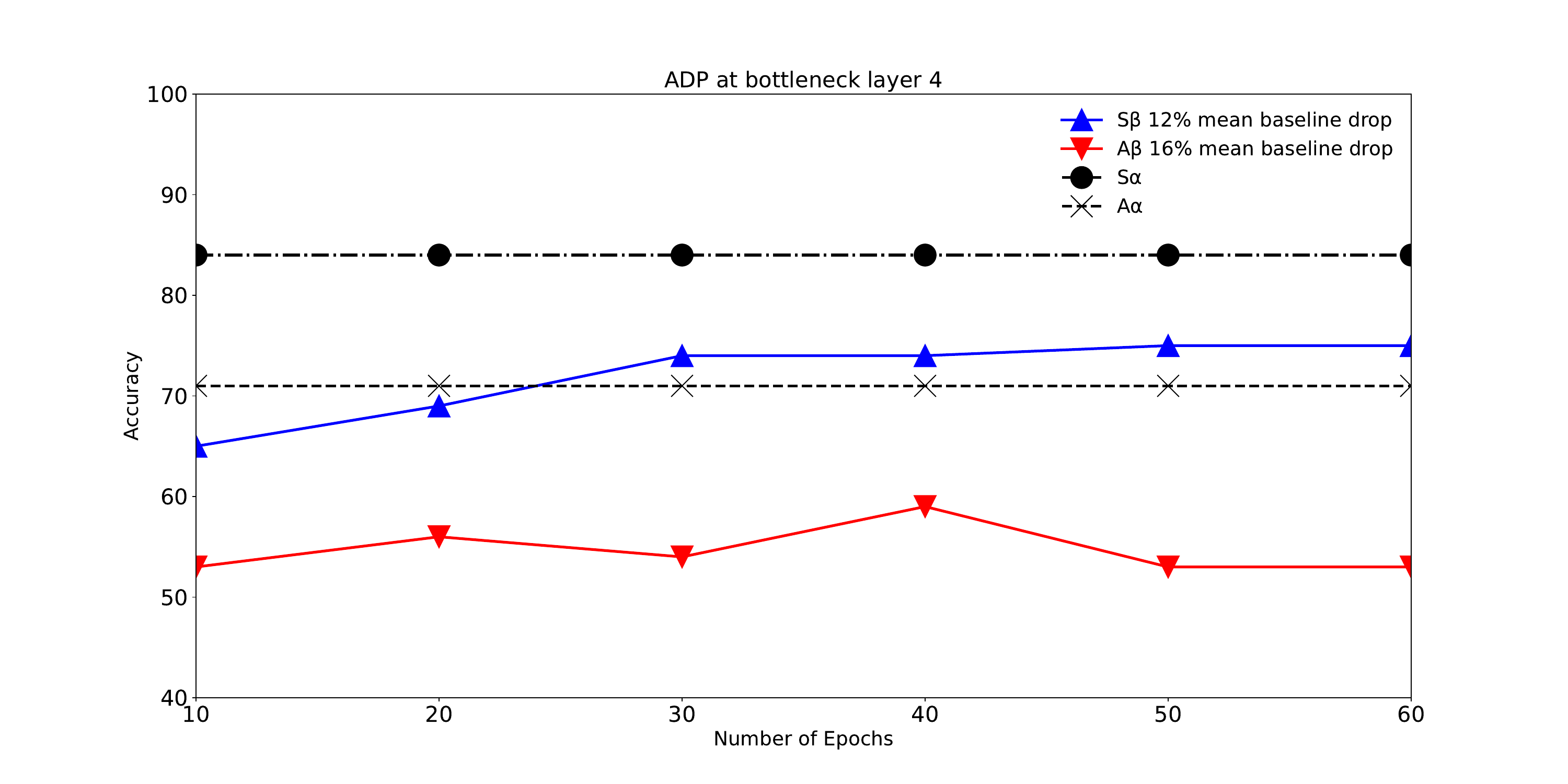}
    \captionsetup{skip=0pt} 
    \caption*{(a)}
  \end{subfigure}
  \hspace{-1em}
  \begin{subfigure}{0.45\textwidth}
    \centering
    \includegraphics[width=\linewidth, trim=50 10 100 50, clip]{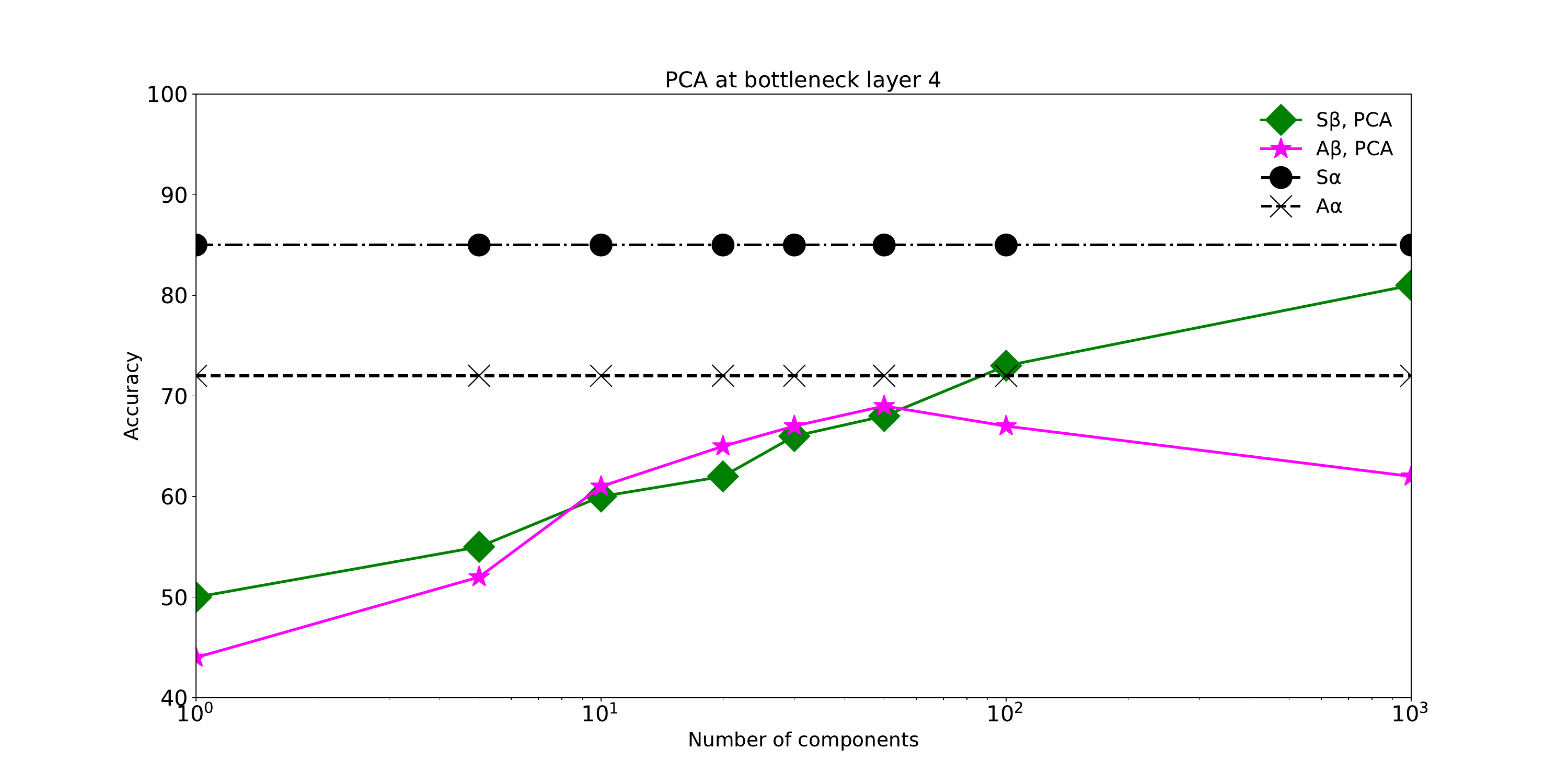}
    \captionsetup{skip=0pt} 
    \caption*{(d)}
  \end{subfigure}
  \hspace{-0.5em}
  \vspace{-0.25em}
  
  \begin{subfigure}{0.45\textwidth}
    \centering
    \includegraphics[width=\linewidth, trim=50 10 100 50, clip]{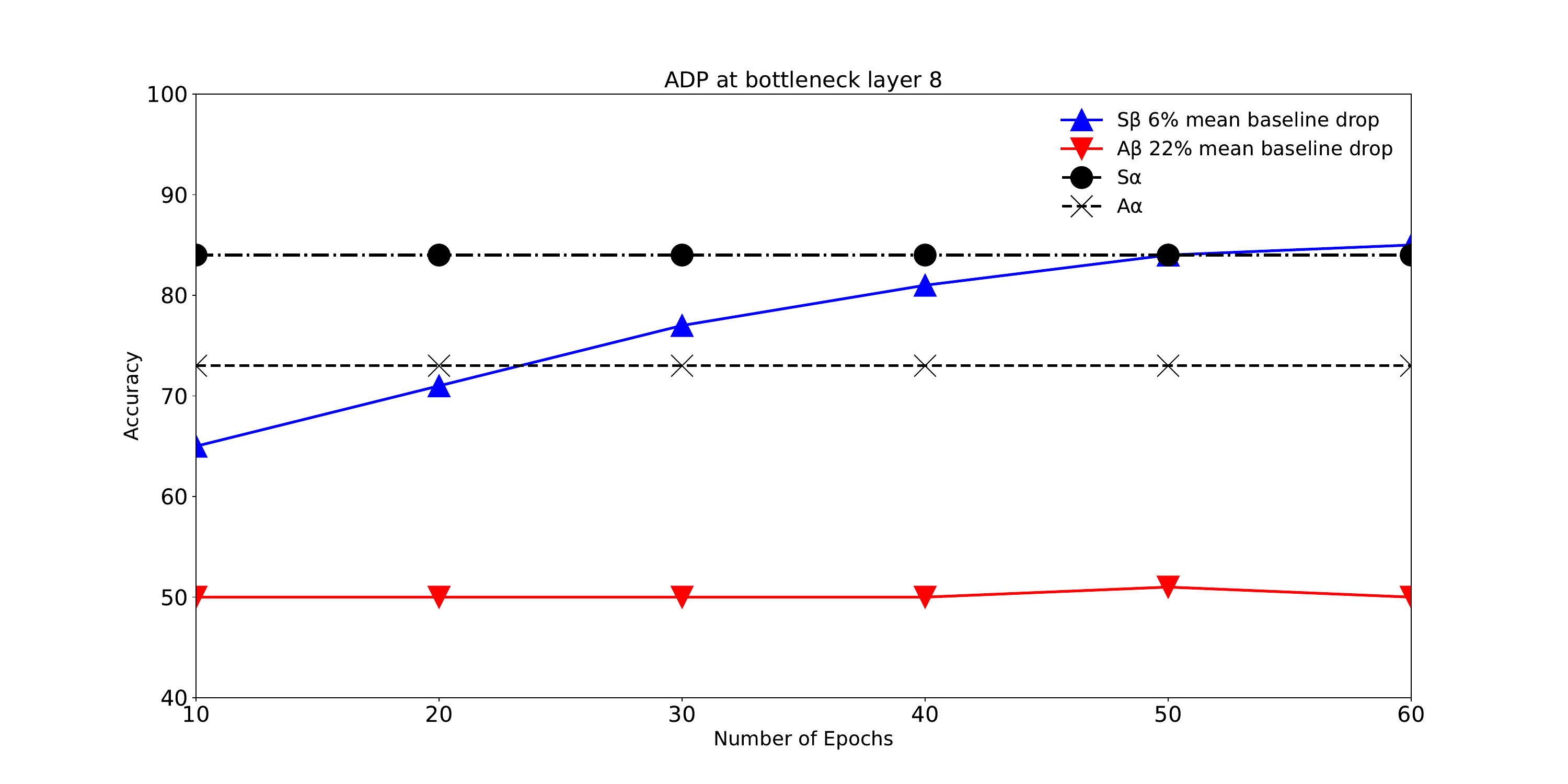}
    \captionsetup{skip=0pt}
    \caption*{(b)}
  \end{subfigure}
  \hspace{-1em}
  \begin{subfigure}{0.45\textwidth}
    \centering
    \includegraphics[width=\linewidth, trim=50 10 100 50, clip]{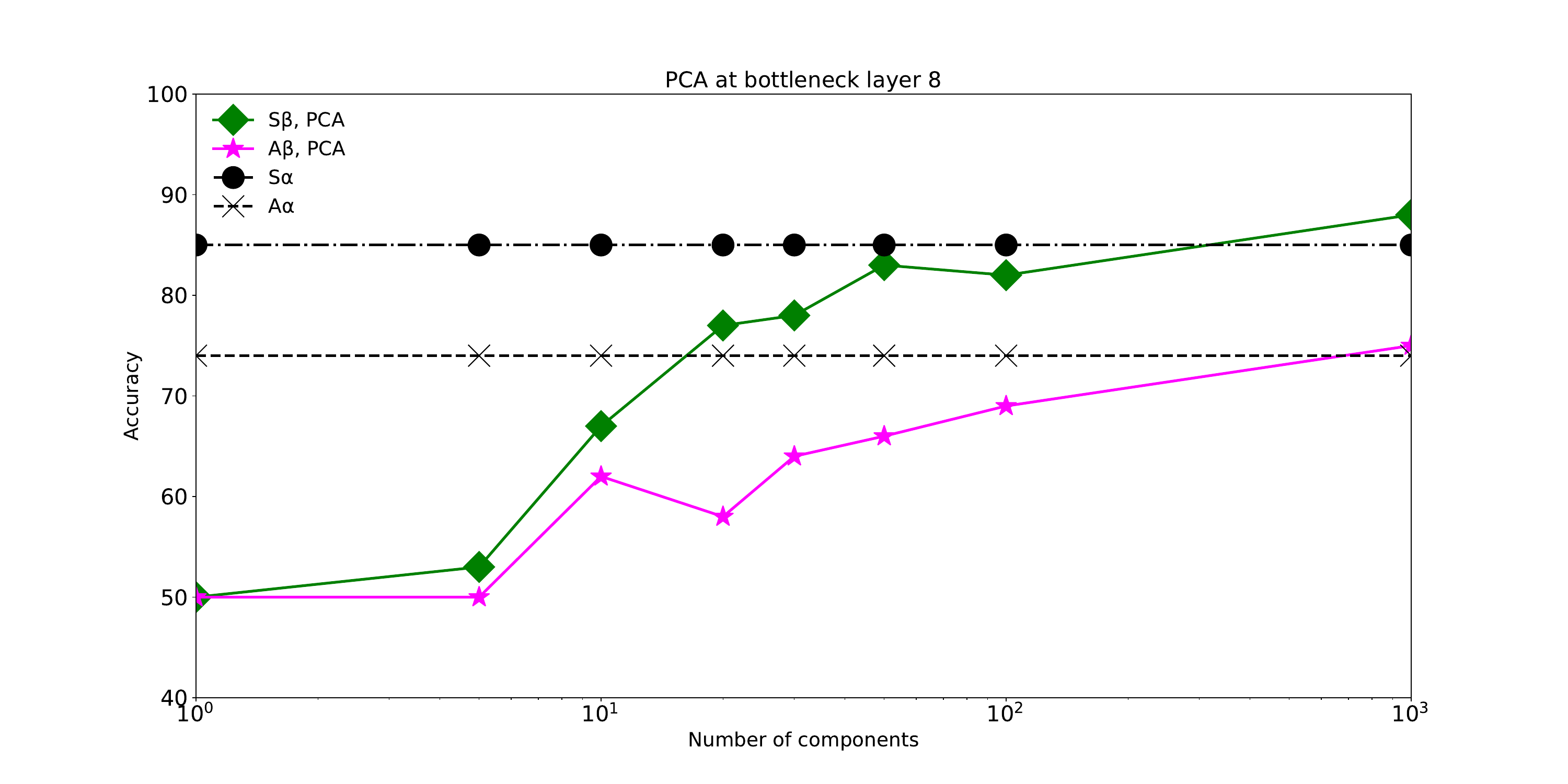}
    \captionsetup{skip=0pt}
    \caption*{(e)}
  \end{subfigure}
  \hspace{-0.5em}
  \vspace{-0.25em}
  
  \begin{subfigure}{0.45\textwidth}
    \centering
    \includegraphics[width=\linewidth, trim=50 10 100 50, clip]{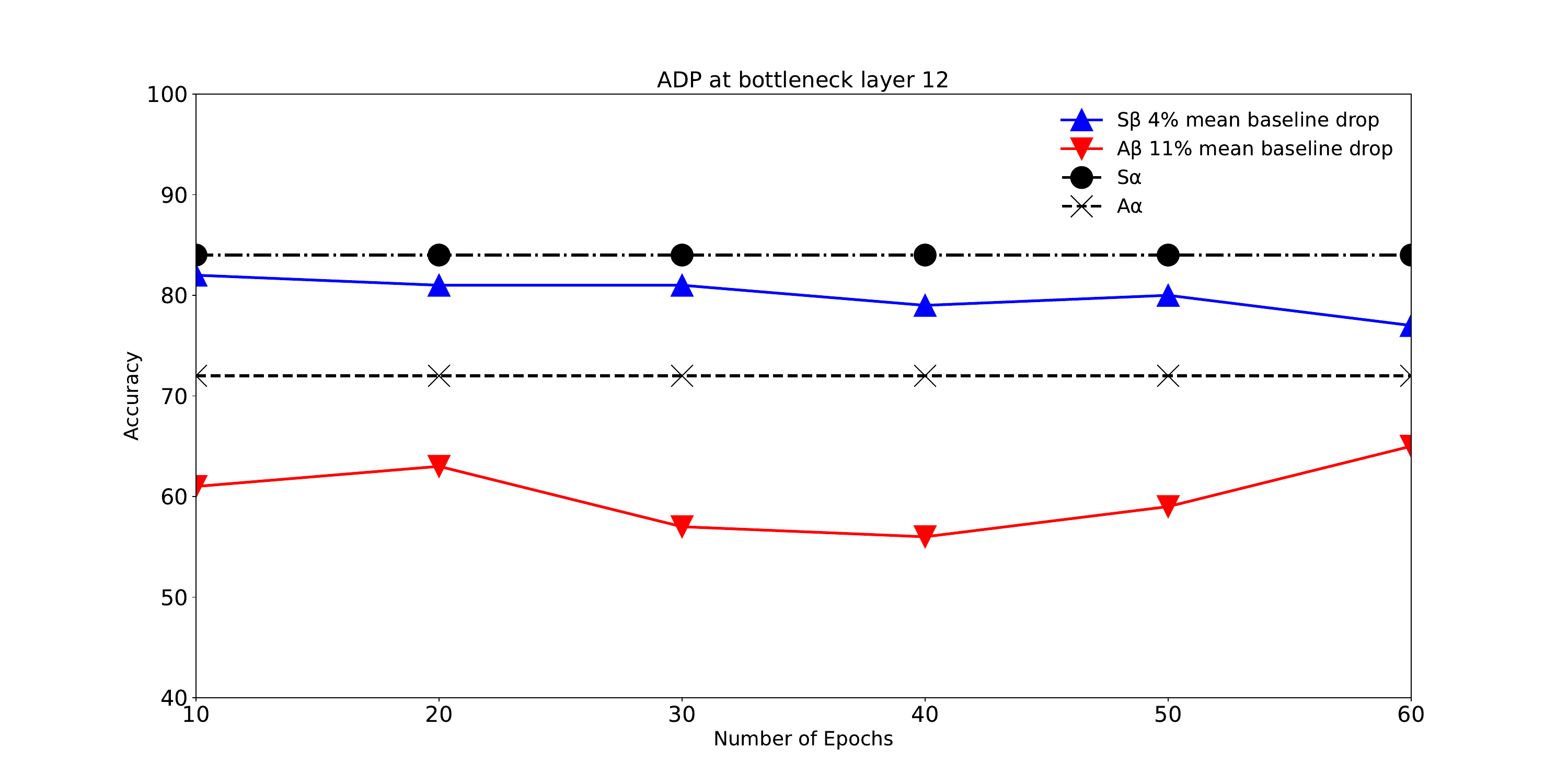}
    \captionsetup{skip=0pt}
    \caption*{(c)}
  \end{subfigure}
  \hspace{-1em}
  \begin{subfigure}{0.45\textwidth}
    \centering
    \includegraphics[width=\linewidth, trim=50 10 100 50, clip]{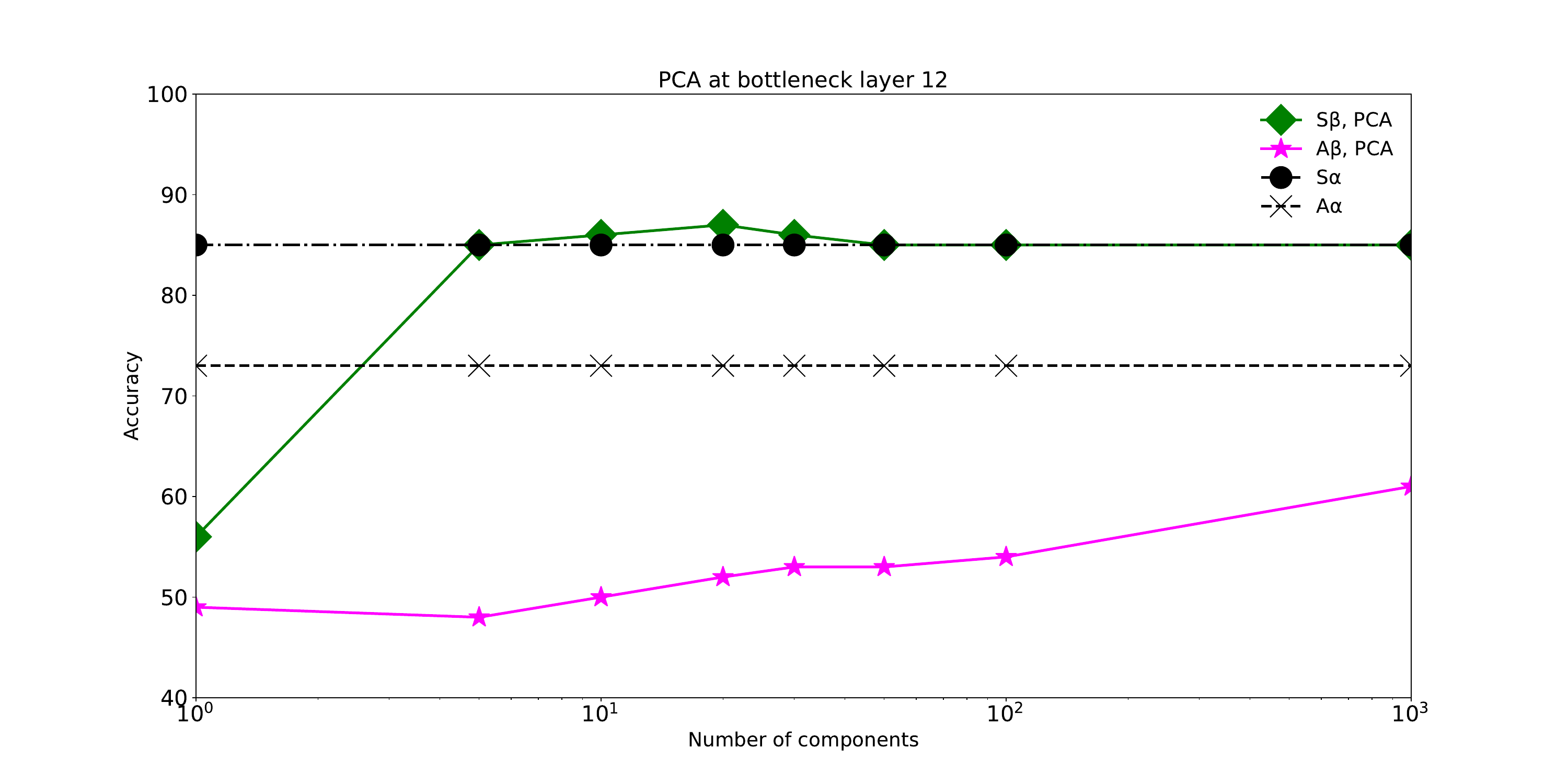}
    \captionsetup{skip=0pt}
    \caption*{(f)}
  \end{subfigure}
  
  \caption{Panel a (b) (c) shows obtained accuracies by $S$ and $A^{i}$ at bottleneck layer 4 (8) (12) with decreasing AE, the ADP strategy: $\Delta_{2}^{min}$ and delta method: black-out, offline adversary architecture: ResNet50 full and inference adversary architecture: ResNet50 split (measured accuracy per epoch). Panel d (e) (f) shows obtained accuracies by $S$ and $A^{i}$ at bottleneck layer 4 (8) (12) with PCA protection. Inference adversary architecture: ResNet50 split (measured accuracy per component).}
  \label{fig:subplot_layout}
\end{figure*}
    
\subsection{Analyzing Results Obtained by ADP}

Based on the results of ADP we observe the following general trends and identify specific promising cases. Our results are reported in Tables \ref{tab:aevgginf}-\ref{tab:aeresinf_deep}, interms of accuracies for the server and adversary before and after protection for each protection strategy as well as the split position the protection was applied to. MS-SSIM values for the backward reconstruction attack are also reported in Table \ref{tab:aevggrec} for VGG16 only. Together these tables provide a view into how the server and adversary are impacted by various ADP strategies as well as their effectiveness at various split positions.

Firstly, in the case of the forward inference attack, we observe that the \textit{$\Delta^{max}$} strategy, shown in rows 5-8 of Table \ref{tab:aeresinf}, rows 5-8 of Table \ref{tab:aeresinf_deep}, and rows 3-4 of Table \ref{tab:aevgginf}, often drops both the server and inference adversary accuracy to a near-random guess. The only exception to this is on row 3 of Table \ref{tab:aevgginf}. However, such a large drop is somewhat expected due to the aggressiveness required by the $\Delta^{max}$ strategy in order to provide protection in the \textbf{extreme cases} that the adversary task is unknown or cannot be inferred. However, even though this strategy is not currently viable it does lend support to the hypothesis that the \textit{blur-out protection method generally allows for higher accuracy results on both the server and the adversary over the black-out method}. Evidence of this can be observed in rows 3-4 of Table \ref{tab:aevgginf}, where the general accuracy drop for the server and the adversary is very pronounced in the split VGG16 case compared to rows 5-6 of Table \ref{tab:aeresinf} in the split ResNet50 case. Additionally, rows 7-8 in Table \ref{tab:aeresinf} and rows 7-8 in Table \ref{tab:aeresinf_deep} exhibit similar trends in different ResNet50 cases. This finding is important because it suggests that a relationship exists between combinations of blurring strength protection and the architectures of the client-server and adversary to produce varying accuracy differentials on the server and adversary. Perhaps even more important, is the implication that even non-important information, indicated by the server CAM cold regions, is still likely needed at some level in order to provide high-level context for the server and the adversary in the final inference stage. 

Secondly, in the case of the $\Delta^{min}$ strategy we generally observe a reduced accuracy drop, shown in rows 1-4 of Table \ref{tab:aeresinf}, rows 1-4 in Table \ref{tab:aeresinf_deep} and rows 1-2 of Table \ref{tab:aevgginf}, compared to the $\Delta^{max}$ strategy. This matches our expectations since the adversary task is known in the $\Delta^{min}$ case thus allowing for it to be specifically targeted. Additionally, the $\Delta^{min}$ strategy also keeps as much unprotected information needed by the server as possible. When comparing the server and inference adversary generally, we found that the accuracy drop observed on the adversary side was also reflected similarly on the server side. Nonetheless, a few promising exceptions exist. Notably in the case of a full-adversary architecture used to train an AE with the black-out method and a split-adversary architecture used to test the protection. In these cases, shown in Figure \ref{fig:subplot_layout}(a-c) and in row 3 of Table \ref{tab:aeresinf}, we found the inference adversary accuracy drops to a near random guess with only a small accuracy drop on the server. 

Thirdly, concerning of the backward reconstruction attack, we define a successful reconstruction attack as one where the attacker achieves an MS-SSIM value exceeding 0.35. Our initial investigation indicated that our best effective attack was only successful against the VGG16 architecture, not the ResNet50 architecture, using the white-box attack described in \cite{he2019model}. This is because the VGG16 architecture is much shallower and simpler lending itself to backward reconstruction attacks more aptly as opposed to the ResNet50 architecture that does not. As such, we only consider the VGG16 architecture, shown in Table \ref{tab:aevggrec} and in Figure \ref{fig:overall}. We observed that in every case we were able to provide protection against the backward reconstruction attack far below the reconstruction threshold. Additionally, we observe that the $\Delta^{max}$ strategy affords slightly better reconstruction protection generally than the $\Delta^{min}$ strategy and so do the black-out and blur-out methods, respectively.

Lastly, by virtue of using a decreasing AE the size of the feature map is reduced in every scenario, where it is applied.

\subsection{PCA Results}

Attained accuracies by adversary and server using PCA at various bottleneck layers are shown in Figure \ref{fig:subplot_layout}(d-f). Thought $A$ and $S$ show the profiles accuracy determined by increasing the number of principal components used in the reconstruction. As the data spans several orders of magnitude, a logarithmic scale was used for plotting the results obtained by PCA to better visualize the data by spreading out the data samples more evenly.

\subsection{Analyzing the Results Obtained by PCA}

The results obtained by the PCA protection indicate that the deeper PCA is integrated into the architecture, the fewer components are needed for reconstruction to achieve good performance. This is characterized by a small drop in the server performance and a significant drop in the adversary performance. Notably, at bottleneck layer 12, using as few as 5 components the server can achieve a performance that matches the server baseline, and the adversary's performance drops to the random guess.

\subsection{Comparing ADP and PCA Outcomes}

\begin{figure}
    \centering 
    \includegraphics[width=\columnwidth, trim=50 10 100 50, clip]{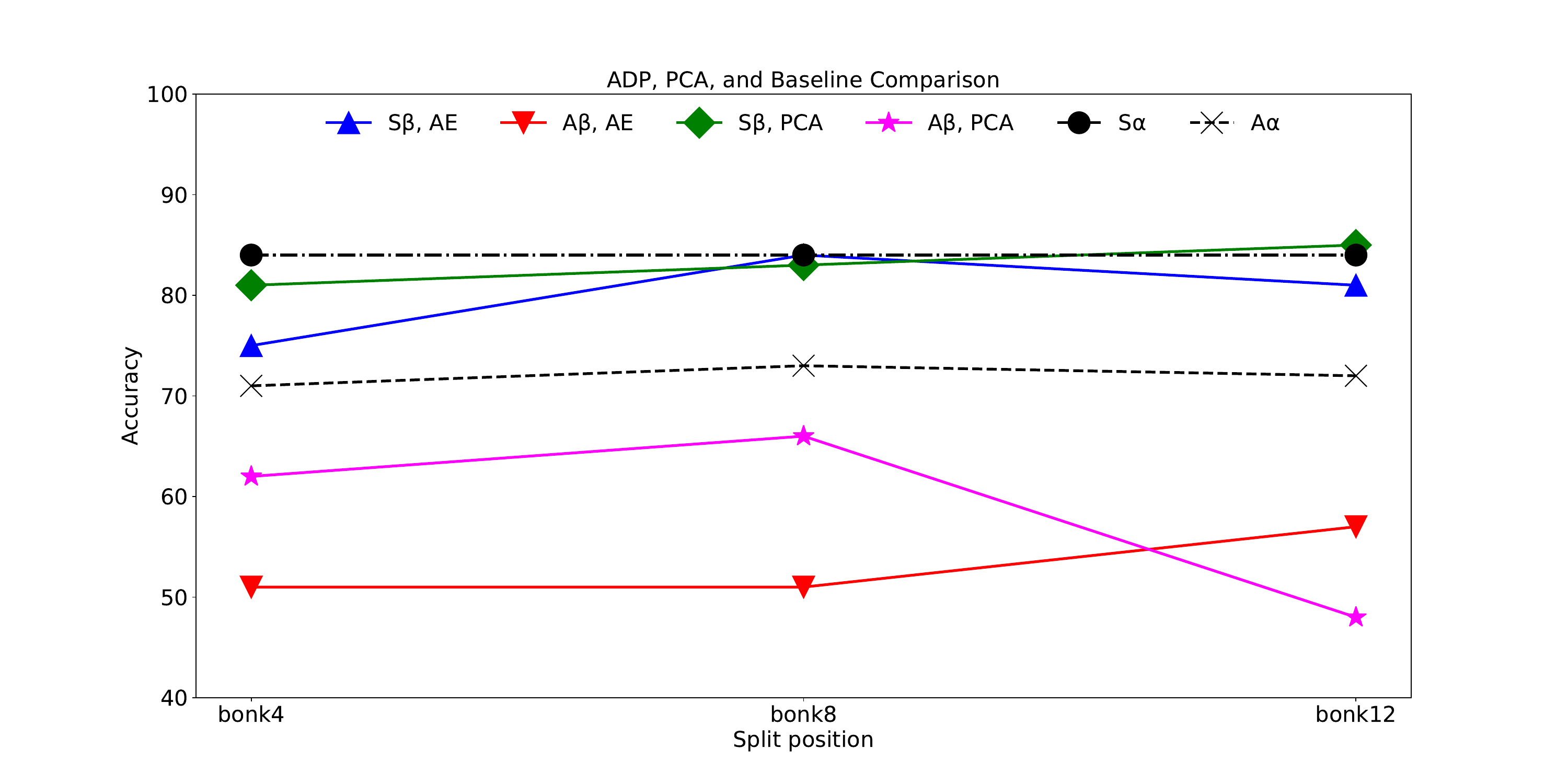}
    \caption{Obtained accuracies by $S$ and $A^{i}$ at bottleneck layers 4, 8, and 12 with ADP and PCA protections. Inference adversary architecture: ResNet50 split.}
    \label{fig_comparison}
\end{figure}

Here, we compare the ADP strategy with the PCA protection on the same primary and sensitive tasks, Wearing\_Lipstick and Wearing\_Hat, respectively, for bottleneck layers 4, 8, and 12 as shown in Figures \ref{fig:subplot_layout} and \ref{fig_comparison}. We also utilize the server and inference adversary weights used in the ADP experiments to ensure results are comparable. Figures \ref{fig:subplot_layout}(a-c) show accuracies attained through ADP as the number of training epochs are increased at bottleneck splits 4, 8, and 12, similarly for Panels (d-f) with respect to PCA with the number of components are increased instead of training epochs. 

Before interpreting the results it is important to reiterate important considerations related to an increase in split position. Firstly, as the split is increased, the client gains more convolutional layers, whereas the server and adversary (in the split adversary case) lose convolutional layers, and, hence, influence over the remaining ``post-protection'' or ``post-split'' decision. However, this is not entirely true for the server as the split position increases, server layers simply shift back onto the client before the protection, presumably making it more consistent with its original baseline model, free from the effects of unwanted perturbation resulting from the applied protection. This is particularly evident in the case of the \textit{split} adversary architecture we analyze here. Consequently, this would suggest a higher initial accuracy for the server as the split position is increased, while the adversary should remain consistently lower. Secondly, regardless of how many convolutional layers remain on the server or adversary, the feature map and information within it will also change and become more compressed (reduced height and width) and varied (greater depth) as the split position is increased, which may impact performance. Lastly, it is important to recognize that the server is disadvantaged compared to any adversary in this case since it is not allowed to retrain on protected feature maps in our setting due to the plugin nature of the protection, while the adversary model can.

At bottleneck layer 4 for ADP, Figure \ref{fig:subplot_layout}(a), we observe that both the server and adversary accuracies remain relatively stable, with a slight increase in server accuracy and a drop in adversary accuracy to a level near random guessing as the number of training epochs increases. This observation generally aligns with Equation \ref{eq1}, Additionally, in the case of PCA, Figure \ref{fig:subplot_layout}(d), it can be observed that as the number of components increases, both server and adversary accuracy increase until they reach a certain point where they begin to diverge. The server accuracy increases to near baseline accuracy, while the adversary accuracy decreases, albeit not reaching the level of a near random guess, thus not satisfying Equation \ref{eq1}. Importantly, at this split position the adversary and the server have the greatest influence on the final decision compared to the client, as they retain most of the convolutional layers.

At bottleneck layer 8 for ADP, Figure \ref{fig:subplot_layout}(b), we observe perfect alignment with Equation \ref{eq1} such that as the number of training epochs increases the server accuracy gradually increases to a point slightly above its previous baseline accuracy, whereas the adversary accuracy remains consistently at a near random guess with little to no deviation. Importantly, PCA results in Figure \ref{fig:subplot_layout}(e), show that as the number of components increases, both the server and adversary converge on their respective baselines ultimately providing little to no protection, even in the best case near $10^{1.25}$ components. At this split position, the adversary and the server have arguably the same influence on the final decision as they both have the same number of convolutional layers as the client.

At bottleneck layer 12 for ADP, Figure \ref{fig:subplot_layout}(c), we observe a very high accuracy, near baseline, for the server that decreases overtime and the adversary accuracy that increases close to its baseline value, as the number of training epochs increases. This trend does not satisfy Equation \ref{eq1}. For PCA, Figure \ref{fig:subplot_layout}(f), we observe that server accuracy sharply rises to its baseline value as more components are added along with the adversary accuracy that increases more gradually. In the best case, near $10^{0.7}$, Equation \ref{eq1}, is fulfilled. At this split position, the adversary and server have arguably limited influence on the final decision as they have fewer convolutional layers compared to the client.

Overall, we observe that the ADP strategy is preferred since it is able to drop the adversary to the near-random guess and maintain an acceptable level of server performance for the bottleneck layers 4 and 8, whereas the PCA plug-in strategy only does so for the bottleneck layer 12, shown in Figure \ref{fig_comparison}. Additionally, ADP is preferred since it can achieve results at an earlier split position where it might be feasibly implemented as opposed to a later split position close to the server. Therefore, we conclude that the ADP strategy is preferred since it can provide protection at an earlier split position with a greater utility at the earliest split for the ResNet50 architecture, whereas PCA does not.

\section{Conclusion}

This work reviews various protection technologies in the existing literature and proposes a novel plug-in-based protection strategy to address both privacy and utility concerns between the server and adversary in the cloud. We compare PCA with our novel method that combines AE and CAMs to implement what we call the ADP approach, which can destabilize the adversary in most if not all cases, while maintaining the server performance with improvements over PCA. Specifically, the plugin nature of the approach and reduced size of the feature map are achieved in all cases by the design and implementation of the ADP approach. Further work is needed to improve the ADP strategy in addition to testing a greater depth and breadth of cases to fully understand how to optimize the protection and achieve the desired universal results.

\printcredits

\bibliographystyle{cas-model2-names}

\bibliography{cas-refs}

\end{document}